\documentclass[a4paper,fleqn]{cas-sc}

\usepackage{amsmath,amssymb,mathtools}
\usepackage{booktabs,tabularx,array,multirow}
\usepackage{tikz}
\usetikzlibrary{arrows.meta,positioning,calc,fit,backgrounds}
\usepackage{stfloats}
\usepackage{hyperref}
\usepackage[c3]{optidef}
\usepackage{amsmath,amssymb,amsfonts,commath,amsthm}
\usepackage{enumitem}
\usepackage{subcaption}
\newtheorem{remark}{Remark}
\usepackage{algorithmic}
\usepackage[ruled, lined, linesnumbered, commentsnumbered, longend]{algorithm2e}
\usepackage{graphicx}
\usepackage{textcomp}
\usepackage{xcolor}
\usepackage{tabularx}
\usepackage{tabulary}
\usepackage{longtable}
\usepackage{verbatim}
\usepackage{setspace}
\usepackage{csvsimple,array,filecontents,booktabs}
\usepackage{multirow}
\usepackage{pdfpages}
\usepackage[acronym]{glossaries}

\usepackage{enumitem}
\usepackage{microtype}
\usepackage{xcolor}
\definecolor{sarblue}{RGB}{42,91,147}
\definecolor{saramber}{RGB}{199,126,40}
\definecolor{sargreen}{RGB}{59,125,90}

\begin{document}
\let\WriteBookmarks\relax
\def\floatpagepagefraction{.80}
\def\textpagefraction{.08}
\shorttitle{Evolutionary-based Transfer Learning for Physical Attack on End-to-End Autoencoder Communications}
\shortauthors{Luong et. al.}
\title[mode=title]{Evolutionary-based Transfer Learning for Physical Attack on End-to-End Autoencoder Communications}
\author[1]{Ho Viet Duc Luong}[orcid = 0000-0001-6690-5222]
\ead{luonghvd@soict.hust.edu.vn}
\author[1]{Lang Hong Nguyet Anh}[orcid = 0009-0001-9975-3357]
\ead{anh.lhn235885@sis.hust.edu.vn}
\author[1]{Huynh Thi Thanh Binh}[orcid = 0000-0003-1976-6113]\ead{binhht@soict.hust.edu.vn}
\author[1]{Trinh~Van~Chien\corref{cor1}}[orcid = 0000-0002-5675-8414]\ead{chientv@soict.hust.edu.vn}
\cortext[cor1]{Corresponding author}
\affiliation[1]{organization={School of Information and Communications Technology, Hanoi University of Science and Technology},
            city={Hanoi},
            postcode={100000},
            country={Vietnam}}

\begin{abstract}
End-to-end autoencoder communication has emerged as a promising means of achieving efficient data transmission across various domains. Nevertheless,  vulnerabilities have recently been revealed in these systems by adversarial attacks, wherein malevolent entities manipulate the transmitted signals to deceive the autoencoder. This paper leverages the applications of evolutionary algorithms to solve a challenging optimization problem of launching physical attacks against end-to-end autoencoder communication. 
For white-box scenarios, we introduce PAGA {(Physical Attack based on Genetic Algorithm)}, which combines a genetic algorithm with elite population initialization to craft attack signals that degrade autoencoder performance. To gauge the efficacy of these attack signals, we employ a fitness function that assesses their capacity to induce substantial reconstruction errors and misclassification rates. A Multi-Factorial version of PAGA, referred to as PAMFGA, is proposed to deal with practical scenarios where the attackers have limited knowledge about the receiver, also known as a black-box attack. Our experimental results show that the proposed attacks can substantially
degrade the reliability of the considered autoencoder communication
systems across the evaluated settings. This study also underscores the critical importance of incorporating security considerations into the design and implementation of end-to-end autoencoder communication systems.
\end{abstract}

\begin{keywords}
Adversarial attacks\sep autoencoder systems\sep wireless security\sep evolutionary algorithms\sep transfer learning
\end{keywords}



\ExplSyntaxOn
\keys_set:nn { stm / mktitle } { nologo }
\ExplSyntaxOff
\maketitle

\section{Introduction}

Deep Learning (DL), a subfield of Artificial Intelligence (AI), focuses on constructing and training Deep Neural Networks (DNNs) to autonomously discover and extract salient representations from data \cite{gao2025comprehensive}. Owing to their multi-layered architectures and capacity to model complex data structures, DL methods have driven substantial progress across diverse domains \cite{dong2021survey}. In computer vision, DL has achieved strong performance in object recognition, image segmentation, facial recognition, and image classification \cite{nani2025enhancing,xia2025comprehensive,anil2022literature,moreira2025deep}. In Natural Language Processing (NLP), DL has advanced machine translation, text classification, summarization, and automatic text generation \cite{gardazi2025bert,koc2025survey}. DL has also demonstrated success in speech-related tasks, including speech recognition, speech synthesis, and voice translation \cite{ivanko2023review}. Collectively, these results highlight DL's ability to learn hierarchical representations that capture intricate dependencies among data attributes.

In parallel, 6G wireless communications (6G) has emerged as a major research frontier, aiming to enable a new era of connectivity with ultra-wide bandwidth in the mmWave/terahertz (THz) spectrum, microsecond-level latency, extremely high reliability, and massive Internet of Things (IoT) connectivity \cite{saad2019vision,song2022terahertz,nguyen20216g}. Several technologies are widely regarded as key enablers. Massive multiple-input multiple-output (MIMO) and its cell-free MIMO variants coordinate multiple access points to perform sophisticated beamforming, thereby improving throughput and coverage \cite{van2016joint,van2021reconfigurable}. Reconfigurable intelligent surfaces (RIS), i.e., metasurface arrays with programmable phase/amplitude responses, can passively ``reconfigure'' propagation conditions to create favorable paths and suppress interference \cite{elmossallamy2020reconfigurable,duc2026evolutionary}. Movable and rotatable antennas add geometric degrees of freedom by enabling antenna repositioning or reorientation to track power hotspots in dynamic environments \cite{zhang2025movable,ning2025movable}. Next-generation sensor networks further contribute via heterogeneous, low-power, self-organizing sensors that support dense spatio-temporal data collection for industrial, agricultural, and healthcare IoT applications \cite{binh2024efficient,luong2025mobile}. Beyond these, heterogeneous networks (HetNets) integrate macro/small cells with satellites, unmanned aerial vehicles (UAVs), and free-space optical links; meanwhile, integrated sensing and communications, mobile edge computing (MEC), and AI-native networking promote cross-layer optimization and autonomous operation \cite{binh2025efficient}. The synergy of these pillars is steering 6G toward an energy-efficient, secure, and flexible ``anywhere--anytime'' network capable of supporting extended reality, collaborative robotics, autonomous vehicles, Industry~5.0, and national-scale digital infrastructure.

Recently, DL has moved beyond conventional Machine Learning (ML) applications
and has been increasingly adopted in wireless communications
\cite{neumann2018learning,meng2025survey}. Wireless systems pose substantial
challenges, including accurate channel modeling, hardware impairments, and
robust protocol design. DL offers a complementary data-driven paradigm for
feature learning and adaptive optimization, with demonstrated applications in
radio resource allocation, channel estimation, power control, and signal
detection
\cite{van2020power,abd2024evolution,rodrigues2021edge}.
{
Beyond these applications, reinforcement learning has also emerged as a
promising paradigm for learning and adapting communication protocols through
interaction with the wireless environment. In particular, recent studies have
investigated multi-agent reinforcement learning for feasibility-aware MAC
protocol learning and for scalable random-access protocol design in massive
machine-type communication networks
\cite{miuccio2024feasible,jadoon2024marl}.
}
A seminal line of work introduced end-to-end DL-based autoencoder communication
systems, where transmitter and receiver are jointly optimized by gradient-based
learning \cite{o2017introduction}.
{
The autoencoder paradigm has also been extended to code-domain
non-orthogonal multiple access schemes such as sparse code multiple access
(SCMA). Existing studies have investigated deep-learning-based SCMA codebook
design and decoding, as well as end-to-end architectures combining
autoencoders and generative adversarial learning to improve communication
performance and robustness
\cite{kim2018scma,wei2021scma,miuccio2022wgan}.
}
Subsequent studies reported improvements in spectral efficiency, energy
management, and robustness under nonlinear distortions
\cite{kang2018deep}, suggesting that DL will be increasingly influential in
the evolution of 6G.
Despite these benefits, integrating DL into communication stacks introduces
new security and reliability vulnerabilities. A central concern is the
susceptibility of DL models to adversarial perturbations: carefully crafted,
often imperceptible modifications that distort internal representations and
induce erroneous outputs
\cite{sadeghi2018adversarial,hoang2024security}. In communication settings,
adversarial perturbations can cause DL-based decoders to reconstruct incorrect
symbols or messages, effectively exploiting learned decision boundaries.
Compared with conventional jamming, which broadly degrades signal quality,
adversarial attacks can selectively mislead decoding with low energy
footprints and minimal spectral signatures. This fragility is particularly
pronounced in end-to-end autoencoder systems, where small channel-input
perturbations may propagate through the encoder--channel--decoder pipeline and
yield disproportionately high BLER at the receiver
\cite{restuccia2020deep,ye2022multiuser,son2024adversarial}. These
observations motivate defenses such as robust training, certified mechanisms,
and monitoring layers tailored to wireless constraints (power, latency, and
hardware non-idealities).


This paper focuses on physical adversarial attacks, where perturbations must survive over-the-air propagation, and considers both white-box and black-box threat models. White-box attackers can exploit full access to the target model to craft perturbations using direct feedback, while black-box attackers operate with limited knowledge and must rely on transferability across surrogate models or query-efficient strategies \cite{papernot2016transferability,son2024adversarial}. To address the non-convexity, stochasticity, and partial observability of wireless environments, we leverage Evolutionary Algorithms (EAs), which are population-based, derivative-free optimizers effective for rugged and multimodal landscapes \cite{li2022evolutionary,binh2023ensemble}. {To this end, the contributions of this paper are summarized as follows.}
\begin{itemize}
\item We formulate an optimization problem that maximizes the error probability under physical adversarial attacks on DL-based wireless communication systems.
\item  For the white-box scenario (full access to decoder architecture and parameters), a novel algorithm named PAGA is introduced. PAGA combines tailored solution representation, elite population initialization and specialized evolutionary operators to search for perturbation vectors that maximize the decoder error while satisfying the energy constraint.
\item  {To address limited-knowledge black-box attacks, we propose PAMFGA, which jointly optimizes perturbations across multiple substitute autoencoders in a shared search space. Its novelty lies in a two-level transfer mechanism: inter-task transfer during evolution and cross-model transfer of an aggregated consensus perturbation to the unknown target autoencoder.}
\item {Numerical results show that PAGA generally achieves stronger attack
performance than the considered baselines, while PAMFGA provides a
favorable attack-strength--runtime trade-off in the intended
multi-surrogate black-box setting.}
\end{itemize}
The rest of this paper is organized as follows: Section~\ref{sec:related-works} outlines the literature review while
Section~\ref{sec:sysmodel} presents the system model of the autoencoder-based communication framework. Section~\ref{sec:white-box-algorithm} and Section~\ref{sec:black-box-algorithm} detail the proposed PAGA and PAMFGA algorithms, respectively. Section~\ref{sec:numerical_results} provides numerical evaluations. Finally, Section~\ref{sec:conclusion_future_work} concludes the paper and outlines future research directions.
\section{Related Works}\label{sec:related-works}
Early work established end-to-end DL-based autoencoder formulations for physical-layer communication, jointly learning transmitter and receiver components within a single optimization pipeline \cite{o2017introduction}.  {
The autoencoder paradigm has subsequently been explored in more specialized
multiple-access settings. For SCMA, a code-domain NOMA technique, learned
encoder--decoder architectures have been used to jointly construct codebooks
and decoding strategies~\cite{kim2018scma}, while deep neural receivers have
been developed to reduce SCMA decoding complexity~\cite{wei2021scma}.
More recently, an end-to-end SCMA architecture combining an autoencoder with
a Wasserstein generative adversarial network has been proposed to improve
robustness to channel noise~\cite{miuccio2022wgan}. These studies further
illustrate the growing use of learned transmitter and receiver components,
whose robustness to intentional physical perturbations is the focus of the
present work.
} Follow-up studies explored DL-driven enhancements for communication objectives such as improved robustness and efficiency, including resilience under nonlinear effects \cite{kang2018deep}. Beyond end-to-end learning, DL has been widely investigated for component-level tasks in wireless systems, including radio resource allocation, channel estimation, power control, and signal detection \cite{van2020power,abd2024evolution,rodrigues2021edge}. These lines of research collectively support the broader premise that DL can serve as a powerful modeling and optimization tool for future 6G systems.

Adversarial threats against DL models have been actively studied in wireless contexts, with evidence that carefully designed perturbations can significantly degrade decoding outcomes in DL-based receivers \cite{sadeghi2018adversarial,sadeghi2019physical,son2024adversarial}. Prior work commonly distinguishes white-box and black-box attacks based on the adversary's knowledge of the target model \cite{papernot2016transferability,son2024adversarial}. Another orthogonal axis is digital versus physical attacks, depending on whether the perturbation is injected directly into the model input or must propagate through the wireless medium \cite{kurakin2016adversarial}. In wireless systems, this distinction is critical: channel fading, synchronization offsets, Doppler effects, noise, and hardware impairments can distort perturbations and reduce the reliability of purely digital designs when deployed over the air \cite{erpek2019deep,sadeghi2019physical}. Consequently, physical adversarial attacks that explicitly account for over-the-air propagation are increasingly viewed as a realistic and impactful threat model, especially for DL-based autoencoder communication pipelines where small input perturbations can yield large BLER increases \cite{restuccia2020deep,ye2022multiuser,son2024adversarial}.
Although adversarial risks in wireless systems have been reported, much of the literature emphasizes digital attacks under idealized assumptions \cite{sadeghi2018adversarial}, which may not translate to realistic radio environments \cite{son2024adversarial}. Meanwhile, effective physical attack synthesis—particularly in black-box scenarios with limited system knowledge—remains comparatively underexplored. This gap motivates transferable, robust, and query-efficient strategies that can operate under noisy, time-varying, and partially observable conditions.

From an algorithmic standpoint, EAs are stochastic, population-based methods inspired by evolutionary principles and are well-suited for nonlinear, multimodal, and derivative-free optimization \cite{li2022evolutionary,binh2023ensemble}. They have been widely applied to communication system optimization problems, including resource allocation, beamforming, user association, and computation offloading in 6G networks \cite{chien2024active,van2024performance,zhao2024joint}. In adversarial signal design, the attack can be framed as a constrained maximization problem: maximize decoding errors subject to energy and feasibility constraints. The resulting loss surfaces are often rugged and may be non-differentiable or unreliable for gradient-based optimization under stochastic channels, making EAs a natural fit.
Among EA variants, GA is particularly attractive for adversarial perturbation construction due to its explicit population diversity maintenance and its crossover/mutation operators, which help balance exploration and exploitation. This can be beneficial for discovering perturbations that remain effective under channel randomness and modulation variability. In contrast to Particle Swarm Optimization (PSO) \cite{kato2025breakout} and Differential Evolution (DE) \cite{bey2025quantum}, GA supports flexible recombination and hybrid initialization strategies that can preserve diversity while improving convergence, making it a strong evolutionary backbone for white-box physical attack synthesis when model feedback is available.
Beyond single-task optimization, MFEA extends EAs to multitasking settings by evolving solutions for multiple related tasks in a unified population and enabling implicit knowledge transfer \cite{dao2021multifactorial}. This is particularly relevant for black-box adversarial attacks: multiple surrogate autoencoder models can be treated as auxiliary tasks, and inter-task transfer can improve the transferability and robustness of evolved perturbations. By exploiting correlations among tasks, the multifactorial framework integrates transfer learning principles into evolutionary optimization, potentially improving cross-model generalization in adversarial contexts.

Compared with the above physical adversarial attack approaches, the novelty
of the proposed methods does not stem from the general use of genetic
algorithms or transfer learning themselves. PAGA specializes evolutionary
optimization to universal physical perturbation design for end-to-end
autoencoder communications by combining a power-constrained perturbation
representation, hybrid random and gradient-informed initialization, and
tailored evolutionary operators for directly maximizing decoding errors.
Existing physical attacks considered in this work, such as FGMA and
RMAEP~\cite{sadeghi2019physical,son2024adversarial}, primarily construct
perturbations through gradient-based optimization, whereas PAGA performs
population-based derivative-free search over the physically constrained
perturbation space.

PAMFGA further differs from conventional surrogate-based black-box transfer
by jointly treating multiple substitute autoencoders as related adversarial
optimization tasks within a single multifactorial population. Its distinction
is therefore the two-level transfer mechanism: adversarial knowledge is first
exchanged among substitute-model tasks during evolutionary optimization, and
the resulting task-wise perturbations are subsequently aggregated into a
consensus direction for transfer to the inaccessible target autoencoder.
Hence, PAMFGA is not presented as a new generic multifactorial evolutionary
algorithm, but as a problem-specific integration of evolutionary multitasking
and cross-model adversarial transfer for physical-layer attacks. Table~\ref{tab:related_attack_comparison} summarizes the main methodological
differences. PAGA differs from the existing gradient-based physical attacks
by directly employing population-based evolutionary search for the
power-constrained perturbation design. PAMFGA further extends this formulation
to the limited-knowledge black-box setting by jointly optimizing multiple
substitute-model tasks and exploiting both inter-task evolutionary transfer
and subsequent cross-model transfer to the unknown target.

\begin{table}[t]
\centering
\caption{Comparison of representative physical adversarial attack approaches
for learning-based wireless communication systems.}
\label{tab:related_attack_comparison}
\small
\resizebox{\columnwidth}{!}{
\begin{tabular}{lccccc}
\toprule
Method & Target system & Optimization & Black-box &
Multi-surrogate & Inter-task transfer \\
\midrule
FGMA~\cite{sadeghi2019physical}
& Autoencoder & Gradient-based & No & No & No \\

RSBA~\cite{sadeghi2019physical}
& Autoencoder & Surrogate-based & Yes & No & No \\

RMAEP~\cite{son2024adversarial}
& RIS--MIMO AE & Gradient-based & No & No & No \\

PAGA (ours)
& Autoencoder & Evolutionary & No & No & No \\

PAMFGA (ours)
& Autoencoder & Multifactorial EA & Yes & Yes & Yes \\
\bottomrule
\end{tabular}}
\end{table}

\begin{figure}[pos=htbp]
\includegraphics[width=\linewidth]{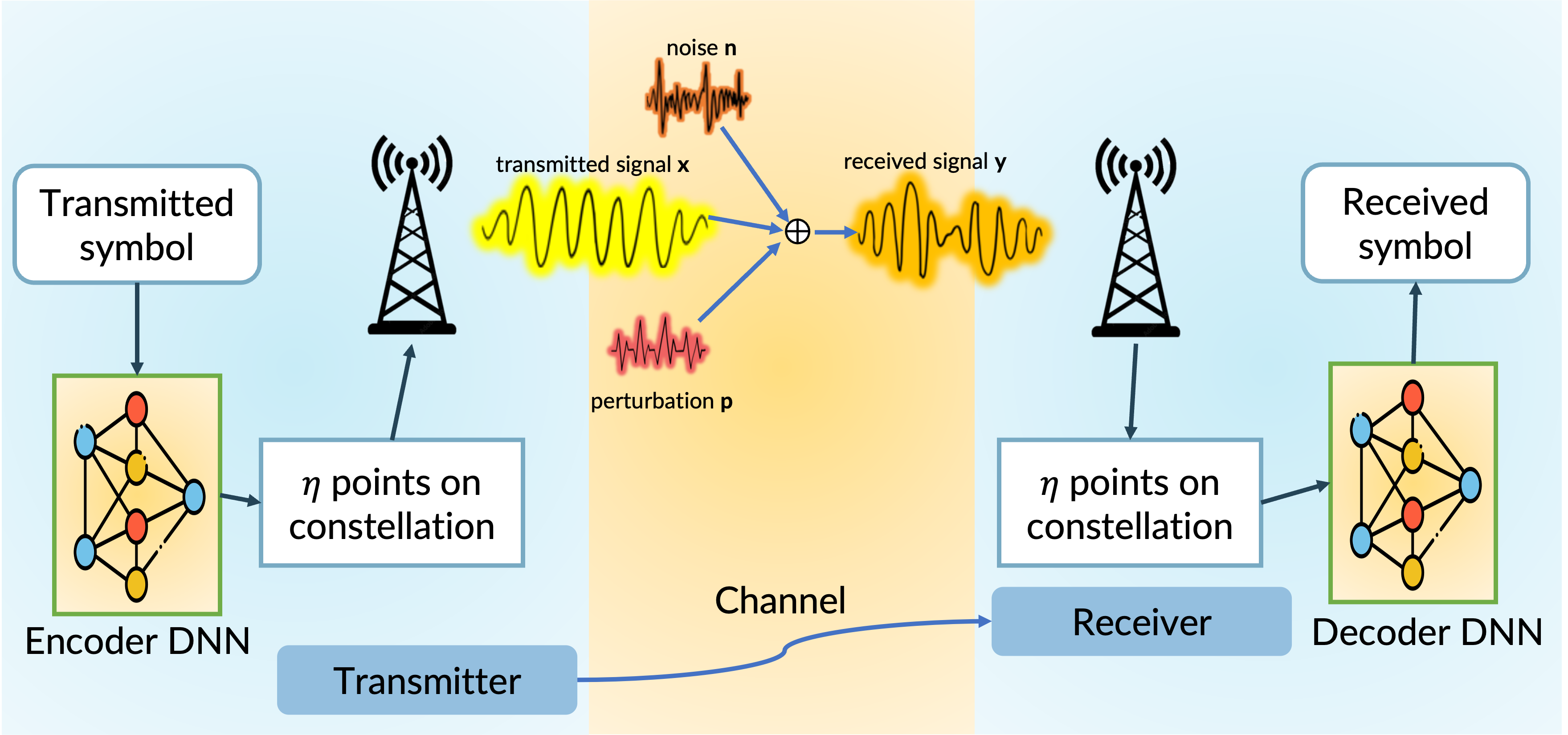}
    \caption{\textbf{An illustration of a wireless communication system based on an autoencoder model.} The encoder DNN at the transmitter maps the input symbol to a constellation of $\eta$ complex-valued points, which are transmitted over a noisy channel. The received signal is corrupted by additive noise $\mathbf{n}$ and adversarial perturbation $\mathbf{p}$, resulting in a modified signal $\mathbf{y}$. The decoder DNN at the receiver processes $\mathbf{y}$ to reconstruct the transmitted symbol. This architecture demonstrates the vulnerability of DL-based communication systems to physical-layer adversarial attacks.
}
    \label{fig:autoencoder}
\end{figure}

\section{System Model, {Preliminaries and} Average Block Error Rate Maximization}\label{sec:sysmodel}

\subsection{System model}
{We consider a point-to-point communication system with three fundamental components comprising a transmitter sending signals to a receiver through a propagation environment. The information to be conveyed is represented by a discrete message symbol
\begin{equation}
    s \in \mathcal{M} = \{1,2,\ldots,M\},
\end{equation}
where the constellation set has cardinality $|\mathcal{M}| = M = 2^{k}$, and $k$ denotes the number of information bits carried by each symbol. It is also assumed that all messages are equiprobable, i.e., 
\begin{equation}
    \mathsf{Pr}(s = m) = \frac{1}{{M}}\ \  \forall  m \in \mathcal{M}.
\end{equation}

At the transmitter, the symbol $s$ is mapped to a complex-valued signal vector through an encoder function
\begin{equation}
    f : \mathcal{M} \rightarrow \mathbb{C}^{\eta},
\end{equation}
where $\eta$ represents the number of channel uses required to send a single symbol. The resulting transmitted signal is denoted by
\begin{equation}
    \mathbf{x} = f(s) \in \mathbb{C}^{\eta}.
\end{equation}
This mapping corresponds to selecting a point in a two-dimensional signal space (per channel use), as in classical modulation schemes such as Quadrature Amplitude Modulation (QAM). To satisfy physical power constraints, the transmitted signal is required to meet an average power constraint of the form
    $\mathbb{E}\bigl[ \lVert \mathbf{x} \rVert_{2}^{2} \bigr] \leq 1$,
where the expectation is taken with respect to the message distribution and any internal randomness of the encoder \cite{chien2024active,chien2024solving}.

The propagation medium is modeled as a quasi-static complex-valued channel impaired by additive noise. In the absence of an adversarial attacker, the received signal can be written as
\begin{equation}
    \mathbf{y} = \mathbf{w} = \mathbf{x} + \mathbf{n},
\end{equation}
where $\mathbf{n} \sim \mathcal{CN}(\mathbf{0}, \sigma^{2} \mathbf{I})$ denotes a circularly symmetric complex Gaussian noise vector with covariance matrix $\sigma^{2} \mathbf{I}$ and noise power $\sigma^{2} = N_{0}/2$. The quantity $\mathbf{w}$ thus represents the \emph{clean} received signal that would be observed in the absence of any intentional interference.

At the receiver, a decoder function
\begin{equation}
    g : \mathbb{C}^{\eta} \rightarrow \mathcal{M}
\end{equation}
is employed to recover an estimate of the transmitted message from the received signal. Given an input vector $\mathbf{y} \in \mathbb{C}^{\eta}$, the decoder outputs an estimated symbol
\begin{equation}
    \hat{s} = g(\mathbf{y}) \in \mathcal{M}.
\end{equation}
From a DL perspective, this straightforward communication system can be regarded as a specialized variant of an \emph{autoencoder} \cite[Chap. 14]{goodfellow2016deep}. Typically, an autoencoder's objective is to identify a low-dimensional representation within an intermediate layer that can faithfully reproduce the output with minimal error. In this manner, the autoencoder acquires the capacity to nonlinearly compress and reconstruct the input data. However, within this particular system, the autoencoder's purpose deviates from the conventional paradigm. Here, the autoencoder seeks to learn representations $\mathbf{x}$ of messages $s$ that exhibit resilience to channel-induced effects, such as noise, attenuation, distortion, hardware flaws, and interference. It also endeavors to acquire the capability to recover transmitted messages from $\mathbf{y}$ with the aim of minimizing error probabilities. In essence, while conventional autoencoders aim to eliminate redundancy from input data for compression, this specialized autoencoder, referred to as the \textbf{channel autoencoder}, often incorporates redundancy. It learns an intermediary representation that can withstand channel-induced transformations.
Fig. \ref{fig:autoencoder} depicts a diagram of an autoencoder system that fulfills the tasks typically associated with classical communication systems. In this representation, the input $s$ undergoes encoding, resulting in a vector. Subsequently, the output is as a probability distribution, contingent upon the signal space. Within this framework, the most probable signal is classified as $\hat{s}$.

The overall communication rate of the system is given by
\begin{equation}
    R = \frac{k}{\eta} \quad \text{[bits/channel use]},
\end{equation}
where $k = \log_{2}(M)$ and $\eta$ is the number of channel uses required to transmit one symbol. Such a system is commonly referred to as an $(\eta,k)$ communication scheme, indicating that one of $M = 2^{k}$ possible messages (equivalently, $k$ bits) is conveyed over $\eta$ channel uses.

\subsection{Physical adversarial attack's mathematical formulation}
\label{subsec:problem_formulation} 


In the presence of an adversarial attacker, a \emph{perturbation} vector
\begin{equation}
    \mathbf{p} \in \mathbb{C}^{\eta}
\end{equation}
is injected into the channel. The resulting received signal at the decoder becomes
\begin{equation}
    \mathbf{y} = \mathbf{w} + \mathbf{p} = \mathbf{x} + \mathbf{n} + \mathbf{p},
\end{equation}
and the decoder produces an estimate
\begin{equation}
    \hat{s} = g(\mathbf{y}) = g(\mathbf{x} + \mathbf{n} + \mathbf{p}).
\end{equation}
The perturbation $\mathbf{p}$ is designed by the attacker with the specific goal of forcing the decoder to output an incorrect symbol, i.e., $\hat{s} \neq s$, while using as little energy as possible or under a fixed energy constraint.

\subsubsection*{Minimum-norm adversarial perturbation for a given symbol}

One way to formalize the attacker's objective is to consider, for a given transmitted symbol $s$ (equivalently, $\mathbf{x}$), the problem of finding the perturbation with minimum $\ell_{2}$-norm that causes a misclassification at the receiver. This can be written as
\begin{equation}\label{eq:min_norm_attack}
	\begin{aligned}
		& \underset{ \mathbf{p} \in \mathbb{C}^{\eta} }{\textrm{minimize}} 
		& & \, \, \lVert \mathbf{p} \rVert_{2} \\
		& \textrm{subject to} && g(\mathbf{w} + \mathbf{p}) \neq f^{-1}(\mathbf{x}),
	\end{aligned}
\end{equation}
where $\mathbf{w} = \mathbf{x} + \mathbf{n}$ and $\mathbf{n} \sim \mathcal{CN}(\mathbf{0}, \sigma^{2} \mathbf{I})$ with $\sigma^{2} = N_0/2$. The constraint $g(\mathbf{w} + \mathbf{p}) \neq f^{-1}(\mathbf{x})$ enforces that the decoder's decision differs from the original message associated with $\mathbf{x}$.

Problem~\eqref{eq:min_norm_attack} corresponds to the classical minimum-norm adversarial example formulation, adapted here to the physical-layer setting. However, this formulation focuses on a single transmitted symbol and does not explicitly capture the average performance degradation over all messages and channel realizations.

\subsubsection*{BLER-based formulation under a power constraint}

To characterize the impact of the perturbation on overall system performance, it is useful to consider the BLER. Assume that the transmitter sends each message in $\mathcal{M}$ with equal probability $1/M$. For a given perturbation vector $\mathbf{p}$, define $\mathsf{BLER}(\mathbf{p})$ as the probability that the received block is decoded incorrectly by the receiver in the presence of $\mathbf{p}$, i.e., the probability that $\hat{s} \neq s$ when $\mathbf{y} = \mathbf{x} + \mathbf{n} + \mathbf{p}$. This probability is taken over the randomness of the transmitted symbol $s$, the noise vector $\mathbf{n}$, and any other stochastic elements of the system.

From the perspective of the attacker, the design challenge can then be formulated as the problem of finding a perturbation $\mathbf{p}$ that maximizes the BLER, subject to an energy constraint on the perturbation itself. This leads to the following optimization problem:
\begin{equation}\label{eq:bler_max_attack}
	\begin{aligned}
		& \underset{ \mathbf{p} \in \mathbb{C}^{\eta} }{\textrm{maximize}} 
		& & \, \, \mathsf{BLER}(\mathbf{p}) \\
		& \textrm{subject to} && \lVert \mathbf{p} \rVert_{2} \leq \mathcal{P}_{\max},
	\end{aligned}
\end{equation}
where $\mathcal{P}_{\max}$ denotes the maximum allowable power available to the attacker, i.e., the upper bound on the energy of the perturbation signal $\mathbf{p}$. The constraint in \eqref{eq:bler_max_attack} ensures that the attack remains within a physically meaningful and practically feasible power budget.

{
Although the perturbation-power constraint in Problem~\eqref{eq:bler_max_attack} defines a simple bounded feasible region, the BLER objective is highly non-convex, non-smooth, and stochastic. The non-convexity originates from the nonlinear composition of the encoder, physical channel, and DNN-based decoder, whereas the discrete decoding decision and the indicator of the error event introduce non-smoothness and possible discontinuities in the objective landscape.
Furthermore, because the received signal depends on random channel noise, the fitness associated with a fixed perturbation is itself a stochastic quantity. Consequently, closed-form optimization and global optimality
guarantees are generally unavailable, while gradient information may be unstable or entirely inaccessible in the black-box setting. In practice, the attacker therefore seeks a high-quality feasible perturbation under a limited evaluation budget rather than an exact global optimum. These characteristics
motivate the use of population-based, derivative-free evolutionary algorithms, which can explore multiple regions of the search space and remain applicable under noisy and partially observable objective evaluations.}

\subsection{White-box and black-box attack settings}
\label{subsec:white_black_settings}

The formulation in \eqref{eq:bler_max_attack} can be instantiated under different assumptions on the attacker's knowledge of the communication system. In this paper, we consider two main scenarios, following the standard taxonomy of adversarial attacks in machine learning: \emph{white-box} and \emph{black-box} settings \cite{papernot2016transferability,sadeghi2019physical,son2024adversarial}.

\subsubsection*{White-box setting}

In a white-box physical adversarial attack, two key assumptions are made regarding the attacker:

\begin{itemize}
    \item The attacker has complete knowledge of the autoencoder-based communication system, including the architecture, weights, and biases of the decoder network $g(\cdot)$. In other words, the internal structure and parameters of the receiver's DNN are fully known.
    \item The attacker is synchronized with the transmitter, so that each element of the perturbation vector $\mathbf{p} \in \mathbb{C}^{\eta}$ (or its real-valued counterpart in $\mathbb{R}^{2\eta}$) is correctly aligned with the corresponding element of the clean signal $\mathbf{w} = \mathbf{x} + \mathbf{n}$.
\end{itemize}
Under these assumptions, the attacker can, in principle, evaluate $g(\cdot)$ and its gradients with respect to the input, or apply any gradient-free optimization strategy to approximate the solution of~\eqref{eq:bler_max_attack}. However, due to the highly non-convex nature of the {DNN} decoder, finding globally optimal perturbations remains a challenging task. Existing works have explored gradient-based approaches, such as the iterative FGMA in \cite{sadeghi2019physical} and the {RMAEP} in \cite{son2024adversarial}, to generate universal perturbations. While these methods provide feasible attack strategies, their solution quality and robustness leave room for improvement.

To address these limitations, this paper later proposes a genetic-algorithm-based framework, referred to as PAGA, which exploits hybrid deterministic and metaheuristic initialization together with tailored evolutionary operators. The white-box scenario thus serves as a setting where full model knowledge can be leveraged to design highly effective physical adversarial attacks.

\begin{remark}[White-box assumptions justification] {
The above white-box assumptions intentionally represent a strong and
controlled attacker model. Similar full-knowledge abstractions are commonly
adopted when evaluating physical adversarial attacks against learned
communication systems, including the related attack settings in
\cite{sadeghi2019physical,son2024adversarial}. Such a setting is useful for
isolating the intrinsic vulnerability of the learned decoder and for
establishing an upper-bound attack scenario before considering uncertainty in
the attacker's knowledge. 
In practical over-the-air deployment, however, perfect synchronization and
perfect channel knowledge may not hold. Timing or carrier-frequency offsets
can misalign the perturbation with the intended received samples, while
channel-estimation errors can alter the amplitude and phase of the
perturbation observed at the receiver. Both effects may reduce the attack
effectiveness obtained under the idealized white-box model. Therefore, the
white-box results in this work should be interpreted as a controlled
full-knowledge benchmark rather than as a claim that perfect synchronization
and channel information are always available to a practical adversary.
}
    
\end{remark}

\subsubsection*{Black-box setting}

In many practical situations, the assumption of full access to the receiver model is unrealistic. {In this paper, we consider a limited-knowledge,
transfer-based black-box setting. The attacker is assumed to know only
system-level information required to construct substitute communication
systems, including the communication configuration $(\eta,k)$ and the
considered physical channel model. However, the attacker has no access to
target-specific information, including the architecture, parameters, weights,
biases, or internal representations of the target encoder and decoder, and
cannot directly query the target model.} In such cases, the attacker must rely on the \emph{transferability} property of adversarial examples: perturbations crafted against one model often remain effective against other models that solve similar tasks, even if their internal structures differ \cite{papernot2016transferability,habler2025adversarial}.
This leads to the black-box adversarial setting in which the attacker designs perturbations using one or more \emph{substitute} models that approximate the behavior of the unknown target decoder. An example of this approach is the RSBA described in \cite{sadeghi2019physical}, where adversarial examples are generated with respect to a surrogate model and then applied to the target system.

To enhance the effectiveness and robustness of physical adversarial attacks in these realistic black-box scenarios, this paper introduces a multifactorial extension of the PAGA algorithm, termed PAMFGA (Multi-Factorial Genetic Algorithm-based Physical Attack). PAMFGA simultaneously optimizes perturbations across multiple surrogate autoencoder models within a unified evolutionary framework, thereby exploiting inter-task transferability to improve attack performance on the unknown target. 
The detailed algorithmic design of PAGA and PAMFGA is presented in the following sections.

\section{Genetic Algorithm-based Adversarial White-box Attack Vector Design}
\label{sec:white-box-algorithm}

Evolutionary algorithms have been widely recognized as powerful tools for solving highly nonlinear, non-convex, and derivative-free optimization problems \cite{huang2021surrogate,ma2025fair}. Motivated by these strengths, this paper adopts a genetic-algorithm-based strategy to design universal physical adversarial perturbations for autoencoder-based communication systems in the white-box setting. The resulting method is referred to as {PAGA}, and its overall workflow is summarized in Fig.~\ref{fig:flowchart-PAGA}.

\begin{figure}[pos=htbp]
    \centering
\includegraphics[width=\linewidth]{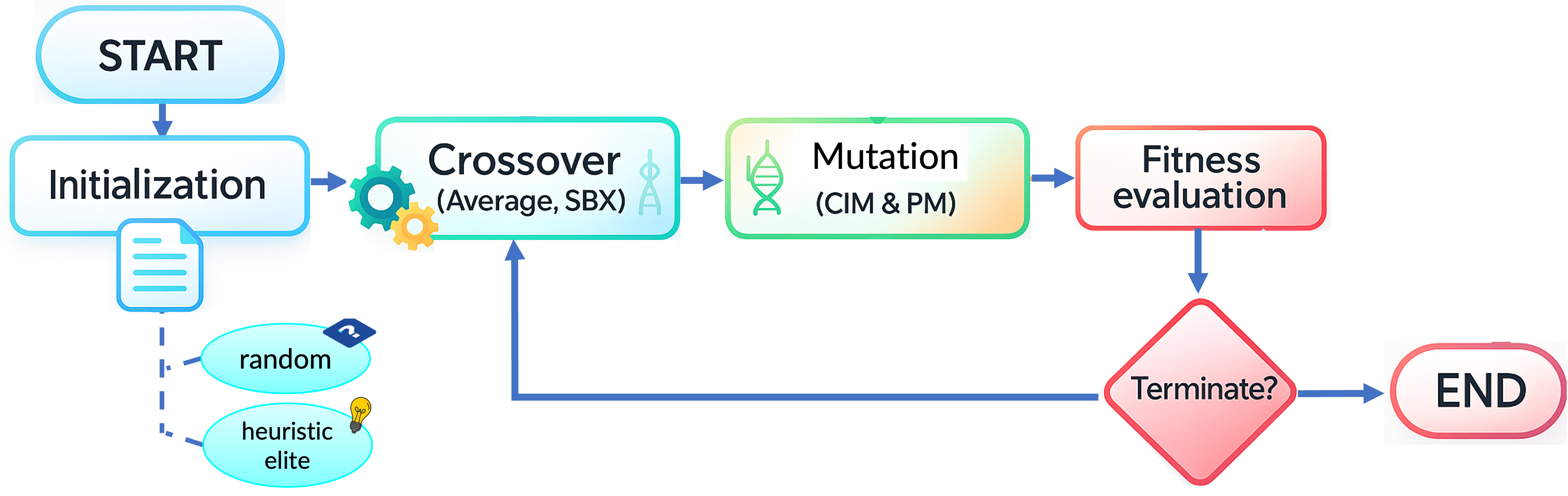}
    \caption{Flowchart of the proposed PAGA framework for the white-box physical attack scenario.}
    \label{fig:flowchart-PAGA}
    \vspace{-.4cm}
\end{figure}

\subsection{Solution representation and population initialization}

In evolutionary algorithms, each individual (or chromosome) encodes a candidate solution of the underlying optimization problem. Within the {PAGA} framework, each individual is associated with a \emph{chromosome} and an \emph{age} attribute. The age records the number of generations during which the individual has remained in the population and is later used by the selection mechanism. The chromosome encodes a perturbation vector $\mathbf{p}$ in real-valued form.

Since the physical perturbation acts on the complex channel output with $\eta$ complex dimensions, each candidate perturbation is represented as a $2\eta$-dimensional real-valued vector, corresponding to the real and imaginary parts of the $\eta$ complex entries. The chromosome is therefore written as
\begin{equation}
    \mathbf{Chro} =
    \begin{bmatrix}
        c_{11} & c_{12} & \cdots & c_{1\eta} \\
        c_{21} & c_{22} & \cdots & c_{2\eta}
    \end{bmatrix}
    =
    \begin{bmatrix}
        \tilde{c}_{1} & \tilde{c}_{2} & \cdots & \tilde{c}_{2\eta}
    \end{bmatrix},
\end{equation}
where each entry $\tilde{c}_i \in \mathbb{R}$ corresponds to one real component of the perturbation. The associated complex perturbation $\mathbf{p} \in \mathbb{C}^{\eta}$ is obtained by grouping consecutive components of $\mathbf{Chro}$ into real and imaginary parts.

To satisfy the power constraint in \eqref{eq:bler_max_attack}, each chromosome must obey
\begin{equation}
    \|\mathbf{Chro}\|_{2}^{2} \leq \mathcal{P}_{\max}.
\end{equation}
If an individual violates this constraint, a normalization step is applied to project it back to the feasible set:
\begin{equation}
    \mathbf{Chro}
    \leftarrow
    \sqrt{\mathcal{P}_{\max}}\,
    \frac{\mathbf{Chro}}{\|\mathbf{Chro}\|_{2}}.
\end{equation}
This operation guarantees that all chromosomes in the population respect the perturbation power budget.

The initial population plays a crucial role in the performance of evolutionary algorithms, as it defines the starting diversity of the search process. {PAGA} employs a hybrid initialization strategy that combines three complementary mechanisms:

\begin{itemize}
    \item \textbf{Uniform random initialization:}  
    Chromosomes are generated by drawing each component independently from a uniform distribution over a pre-defined interval and then applying the normalization step. This strategy promotes broad exploration of the search space at the beginning of the algorithm.

    \item \textbf{Normal random initialization:}  
    Chromosomes are generated from a circularly symmetric complex Gaussian distribution that inherently complies with the perturbation power budget. In practice, the real-valued chromosome entries are sampled from a normal distribution and then normalized such that $\|\mathbf{Chro}\|_{2}^{2} = \mathcal{P}_{\max}$, which emulates $\mathbf{Chro} \sim \mathcal{CN}(\mathbf{0}, \mathcal{P}_{\max})$.

    \item \textbf{Elite (FGM-based) initialization:}  
    To inject problem-specific information into the initial population, a small subset of individuals is generated by first selecting $m$ symbols uniformly at random from the constellation set $\mathcal{M}$ and then computing perturbations $\mathbf{p}$ that disrupt these symbols using a fast-gradient-based method inspired by \eqref{eq:min_norm_attack} and \cite{naqvi2023adversarial}. The resulting perturbations are encoded into chromosomes and added to the initial population as ``elite'' seeds.
\end{itemize}

These complementary initialization strategies provide a mixture of purely random exploration and informed starting points, helping {PAGA} avoid poor initial diversity and accelerating convergence toward high-quality adversarial perturbations.

\subsection{Evolutionary operators: crossover and mutation}

Evolutionary operators are responsible for exploring and exploiting the search space by recombining and perturbing chromosomes across generations. In {PAGA}, two types of crossover operators and two types of mutation operators are employed to balance exploration, exploitation, and population diversity. Because all chromosomes are normalized to satisfy the power constraint, each gene $\tilde{c}_i$ is restricted to a bounded interval, which is taken as $LB = -1$ and $UB = 1$ for notational convenience.

\subsubsection{Hybrid and simulated binary crossover}

The first recombination operator is a {HX} that combines point-wise inheritance and averaging. HX aims to preserve promising structures from the parents while also producing novel combinations of genes. Consider two parent chromosomes
\begin{equation}
    \mathbf{par}^{1} = [p^{1}_{1}, \ldots, p^{1}_{2\eta}], \qquad
    \mathbf{par}^{2} = [p^{2}_{1}, \ldots, p^{2}_{2\eta}],
\end{equation}
and an offspring chromosome
\begin{equation}
    \mathbf{off} = [o_{1}, o_{2}, \ldots, o_{2\eta}].
\end{equation}
Two integer cut-points are drawn at random:
\begin{equation}
    \mu^{1} \sim \mathcal{U}\bigl(0, \lfloor 2\eta/3 \rfloor\bigr), 
    \qquad
    \mu^{2} \sim \mathcal{U}(\mu^{1}, 2\eta),
\end{equation}
which divide the chromosome into three segments: $[1,\mu^{1})$, $[\mu^{1},\mu^{2})$, and $[\mu^{2},2\eta]$. The offspring genes are then defined as
\begin{equation}
    o_{i} =
    \begin{cases}
        p^{1}_{i},                           & i \in [1,\mu^{1}), \\
        \dfrac{p^{1}_{i} + p^{2}_{i}}{2},   & i \in [\mu^{1},\mu^{2}), \\
        p^{2}_{i},                           & i \in [\mu^{2},2\eta].
    \end{cases}
\end{equation}
Shorter middle segments lead to offspring closer to the parents (strong exploitation), whereas longer middle segments promote more extensive mixing of parental information (enhanced exploration).

To further enrich the search dynamics, {PAGA} also adopts the simulated binary crossover (SBX), a widely used real-coded crossover operator \cite{chacon2018analysis,van2024performance}. SBX aims to mimic the spread of offspring in standard binary crossover while operating directly on real-valued genes. For each gene index $i \in \{1,\ldots,2\eta\}$, assume without loss of generality that $p^{1}_{i} \leq p^{2}_{i}$. Given bounds $LB$ and $UB$, the following auxiliary quantities are computed:
\begin{equation}
    \nu^{1}_{i} = 1 + \frac{2(p^{1}_{i} - LB)}{p^{2}_{i} - p^{1}_{i}},
    \qquad
    \nu^{2}_{i} = 1 + \frac{2(UB - p^{2}_{i})}{p^{2}_{i} - p^{1}_{i}}.
\end{equation}
Let $\eta_{c} \geq 2$ denote the SBX distribution index and define
\begin{equation}
    \vartheta^{j}_{i} = 1 - \left(\nu^{j}_{i}\right)^{-(\eta_{c}+1)}, 
    \qquad j \in \{1,2\}.
\end{equation}
Drawing $u_{i} \sim \mathcal{U}[0,1]$, the spread factors $\beta^{j}_{i}$ are given by
\begin{equation}
    \beta^{j}_{i} =
    \begin{cases}
        \bigl(2\vartheta^{j}_{i} u_{i}\bigr)^{\frac{1}{\eta_{c}+1}}, 
        & u_{i} \leq \dfrac{1}{2\vartheta^{j}_{i}}, \\[0.3em]
        \bigl(2 - 2\vartheta^{j}_{i} u_{i}\bigr)^{\frac{1}{\eta_{c}+1}}, 
        & \text{otherwise},
    \end{cases}
\end{equation}
and the two offspring genes are finally computed as
\begin{equation}
    o^{j}_{i}
    = \frac{1}{2}\Bigl[ (p^{1}_{i} + p^{2}_{i}) 
    + \beta^{j}_{i}(p^{1}_{i} - p^{2}_{i}) \Bigr], 
    \qquad j \in \{1,2\}.
\end{equation}
Larger values of $\eta_{c}$ produce offspring that lie close to their parents (favoring exploitation), whereas smaller values permit offspring to spread more widely in the search space (favoring exploration). In {PAGA}, HX and SBX are applied with pre-specified probabilities, and the resulting offspring are subsequently normalized to satisfy the perturbation power constraint.

\subsubsection{Central inversion and polynomial mutation}

Mutation operators introduce stochastic variations into chromosomes, enhancing diversity and reducing the risk of premature convergence to poor local optima. {PAGA} integrates two complementary mutation operators: central inversion mutation (CIM) and polynomial mutation (PM).

For a given chromosome
\begin{equation}
    \mathbf{Chro} = [\tilde{c}_{1}, \tilde{c}_{2}, \ldots, \tilde{c}_{2\eta}],
\end{equation}
CIM operates by reversing the order of genes around a randomly chosen pivot. Specifically, a mutation index $k \in \{1,\ldots,2\eta\}$ is drawn uniformly at random, and the mutated chromosome is defined as
\begin{equation}
    \mathbf{Chro}' =
    [\tilde{c}_{k}, \tilde{c}_{k-1}, \ldots, \tilde{c}_{1},
     \tilde{c}_{2\eta}, \tilde{c}_{2\eta-1}, \ldots, \tilde{c}_{k+1} ].
\end{equation}
This operation preserves the gene values but rearranges their positions, potentially uncovering new structural patterns in the perturbation vector.

Polynomial mutation modifies gene values according to a polynomial probability distribution that respects the variable bounds $[LB,UB]$ and is controlled by a distribution parameter $\eta_{m} > 0$. For each gene $\tilde{c}_{i}$, define
\begin{equation}
    \delta_{i}
    = \frac{\min\bigl(\tilde{c}_{i} - LB,\, UB - \tilde{c}_{i}\bigr)}{UB - LB},
\end{equation}
and draw $\mu \sim \mathcal{U}(0,1)$. The mutation step $\bar{\delta}_{i}$ is then given by
\begin{equation}
    \bar{\delta}_{i} =
    \begin{cases}
        \bigl[2\mu + (1 - 2\mu)(1 - \delta_{i})^{\eta_{m}+1}\bigr]^{\frac{1}{\eta_{m}+1}} - 1, 
        & \mu \leq 0.5, \\[0.3em]
        1 - \bigl[2 - 2\mu + (2\mu - 1)(1 - \delta_{i})^{\eta_{m}+1}\bigr]^{\frac{1}{\eta_{m}+1}},
        & \text{otherwise}.
    \end{cases}
\end{equation}
Collecting all $\bar{\delta}_{i}$ into a vector
\begin{equation}
    \boldsymbol{\Delta}
    = [\bar{\delta}_{1}, \bar{\delta}_{2}, \ldots, \bar{\delta}_{2\eta}],
\end{equation}
the mutated chromosome is computed as
\begin{equation}
    \mathbf{Chro}' = \mathbf{Chro} + \boldsymbol{\Delta}\,(UB - LB).
\end{equation}
The parameter $\eta_{m}$ controls the magnitude of perturbations: larger values yield small, fine-grained mutations around the current gene values, while smaller values allow for larger jumps in the search space. After mutation, the chromosome is again normalized to ensure that the power constraint is satisfied.

\subsection{Selection and termination condition}

In the considered setting, evaluating the quality of a candidate perturbation is computationally intensive, as it requires simulating the autoencoder system, transmitting a set of messages through the channel, and counting the number of incorrectly decoded blocks. Moreover, the objective function is inherently stochastic due to channel noise.

To obtain a stable fitness estimate with limited computational cost, the evaluation in {PAGA} assumes an idealized data sequence in which the transmitter sends each symbol in the constellation set $\mathcal{M}$ with equal probability $\mathsf{Pr}(s = i) = 1/M$ for all $i \in \mathcal{M}$. For a given chromosome (perturbation), only $M$ signals corresponding to all symbols in $\mathcal{M}$ are simulated. The \emph{fitness} of an individual is then defined as the proportion of these $M$ signals for which the presence of the perturbation leads to a decoding error at the receiver. This fitness measure approximates the BLER associated with the perturbation while keeping the simulation complexity manageable. Algorithm \ref{alg:BLER} provides the pseudo code for {PAGA}'s fitness evaluation.

\begin{algorithm}[t]
\caption{Fitness Evaluation (Approximate BLER)}
\label{alg:BLER}
\begin{algorithmic}[1]
\REQUIRE Individual \text{ind}, autoencoder $(f,g)$, constellation $\mathcal{M}$ with $|\mathcal{M}| = M$, noise variance $\sigma^2$.
\ENSURE Updated fitness value $\text{ind.fitness}$.

\STATE Decode perturbation $\mathbf{p}$ from $\text{ind.Chro}$.
\STATE $\text{err} \gets 0$.
\FORALL{$s \in \mathcal{M}$}
  \STATE Compute transmitted signal $\mathbf{x} = f(s)$.
  \STATE Draw noise $\mathbf{n} \sim \mathcal{CN}(\mathbf{0}, \sigma^2 \mathbf{I})$.
  \STATE Form received signal $\mathbf{y} = \mathbf{x} + \mathbf{n} + \mathbf{p}$.
  \STATE Compute decoded symbol $\hat{s} = g(\mathbf{y})$.
  \IF{$\hat{s} \neq s$}
    \STATE $\text{err} \gets \text{err} + 1$.
  \ENDIF
\ENDFOR
\STATE $\text{ind.fitness} \gets \text{err} / M$ \quad \textit{// approximate BLER}
\end{algorithmic}
\end{algorithm}

\begin{remark}[Stochastic fitness approximation and universal perturbation]
The exact BLER is defined as an expectation over both the transmitted-message
distribution and the channel-noise distribution. During evolutionary
optimization, Algorithm~1 uses a computationally inexpensive stochastic
estimate of this objective. In the default setting, all $M$ equiprobable
messages are evaluated, while one independently sampled channel-noise
realization is used for each message. Therefore, the message distribution is
enumerated exactly, whereas the noise expectation is approximated
stochastically.
Because this stochastic estimate directly affects evolutionary selection and
elitism, a small number of noise realizations may produce an optimistic
in-search fitness value for some candidate perturbations. To quantify this
effect, Section~6.2 provides an additional sensitivity analysis with
$N_{\mathrm{fit}}\in\{1,5,10\}$ noise realizations per message. 
The term ``universal perturbation'' refers to the fact that the same
perturbation vector is applied to all transmitted messages. It does not imply
that the perturbation must cause a decoding error for every instantaneous
channel-noise realization.
\end{remark}

After fitness evaluation, a selection mechanism is applied to control population size and preserve high-quality individuals. First, individuals are sorted in descending order of fitness. Age-based elimination is then performed: individuals whose age exceeds a predefined maximum and are not part of the elite set are removed. The number of elite individuals retained at each generation is governed by a hyperparameter $p_{e}$, which helps maintain the best chromosomes found so far. Next, among the remaining individuals, those with worse fitness are iteratively removed until the population size is reduced to the prescribed value $\mathsf{Pop_{size}}$. 

The evolutionary search of {PAGA} proceeds for a fixed maximum number of generations $G_{\max}$. When this termination condition is met, the algorithm outputs the best individual in the final population, which corresponds to the universal physical perturbation estimated to maximize the BLER under the given power constraint in the white-box setting.

\section{Multifactorial Algorithm for Black-box Attack Vector Design}
\label{sec:black-box-algorithm}

{This section presents the PAMFGA framework for the more practical limited-knowledge, transfer-based black-box attack setting. The adversary is assumed to know the system-level communication configuration $(\eta,k)$ and the considered physical channel model, which are used to construct and train substitute autoencoders. However, the adversary has no access to target-specific information, including the architectures, parameters, weights, biases, or internal representations of the encoder and decoder of the target
autoencoder system, and cannot directly query the target model.} The central idea is to exploit the transferability of adversarial perturbations: adversarial examples crafted for one model often remain effective against other models solving the same task with high probability. {PAMFGA adopts established multifactorial evolutionary concepts, including a
unified search space, skill factors, scalar fitness, and inter-task mating. Its novelty lies in adapting these concepts to limited-knowledge physical adversarial attacks and extending the conventional multitask workflow with a second transfer stage toward an inaccessible target model. Specifically, multiple substitute autoencoders are modeled as related attack tasks that share the same perturbation representation and power constraint. During optimization, adversarial information is transferred among these surrogate tasks. After optimization, the task-wise elite perturbations are further aggregated into a consensus adversarial direction and transferred to the unknown target autoencoder.} The overall procedure is depicted in Fig.~\ref{fig:PAMFGA} of {PAMFGA}.

\begin{figure}[pos=htbp]
    \centering
    \includegraphics[width=\linewidth]{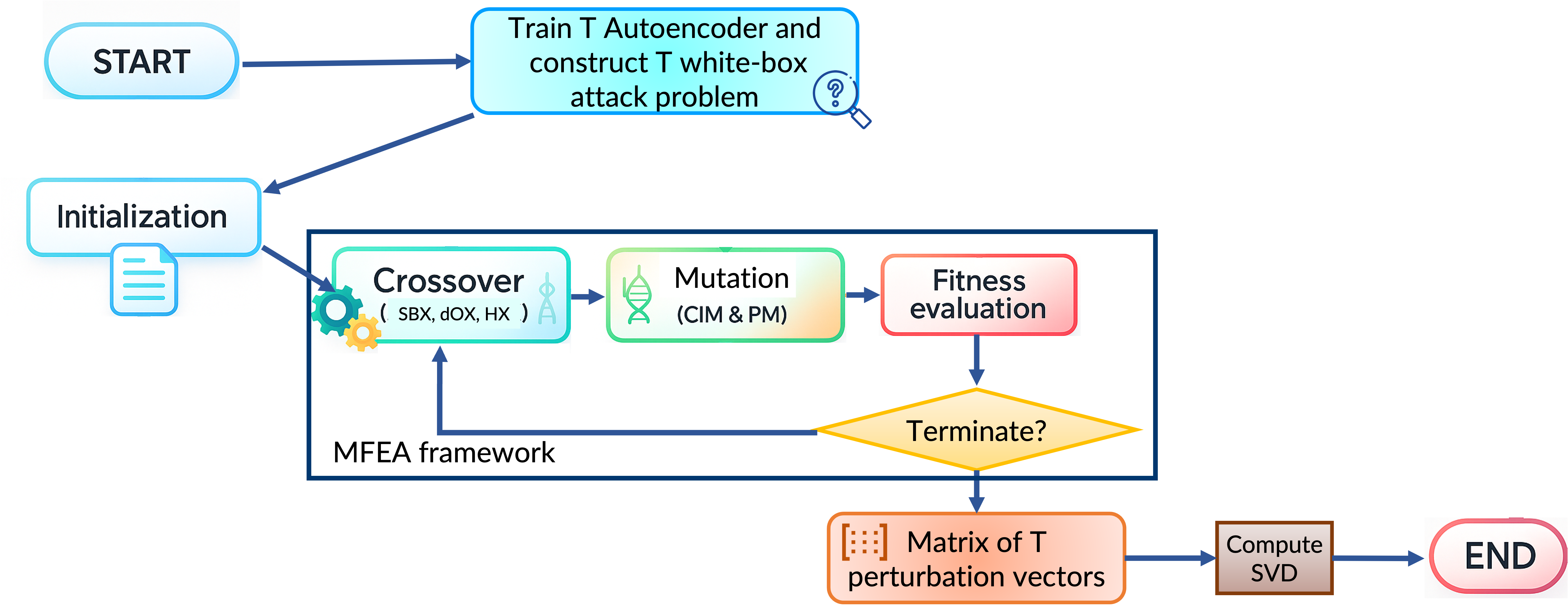}
    \caption{{Flowchart of the proposed PAMFGA framework for the limited-knowledge black-box physical attack scenario, where the attacker knows the system-level communication configuration but has no access to the architecture or parameters of the target autoencoder.}}
    \label{fig:PAMFGA}
    \vspace{-0.5cm}
\end{figure}

\subsection{Population initialization in the unified search space}

Population initialization plays a crucial role in {PAMFGA}, as it directly affects both the exploratory capability of the algorithm and the quality of inter-task knowledge transfer. In the black-box setting, the attacker is assumed to know only the overall system configuration, specifically the pair $(\eta,k)$, while the internal structures of the true encoder and decoder remain unknown.

To approximate the unknown autoencoder, $T$ substitute autoencoder models are constructed and trained offline to solve the same end-to-end communication task with configuration $(\eta,k)$. Each substitute autoencoder is trained to minimize its reconstruction or classification error, thereby mimicking the behavior of a possible target model. This construction leads naturally to a multitasking formulation consisting of $T$ optimization tasks, where
\begin{itemize}
    \item task $t$ corresponds to designing a universal physical perturbation that degrades the performance of substitute autoencoder $t$, and
    \item all tasks share the same perturbation search space $\mathbb{C}^{\eta}$ (or its $2\eta$-dimensional real-valued representation).
\end{itemize}

Within the multifactorial evolutionary framework, these tasks are embedded into a {USS} whose dimensionality coincides with the common dimensionality of the perturbation vectors. The USS enables chromosomes to be interpreted consistently across all tasks, thereby allowing genetic material (candidate perturbations) to be exchanged among tasks without additional encoding or decoding operations. This unified representation is essential for effective multitask optimization and transfer learning.

The chromosome representation in {PAMFGA} is identical to that used in {PAGA}: each individual encodes a $2\eta$-dimensional real-valued perturbation vector and is associated with an age attribute. In addition, every individual is assigned a \emph{skill factor} $\tau \in \{1,\ldots,T\}$ indicating the task for which it is primarily responsible.
Population initialization follows the same three-component strategy as in {PAGA} (uniform random, normal random, and gradient-based/elite initialization), but adapted to the multitask context. In particular, for each substitute autoencoder (task) $t$, one elite perturbation is generated using a white-box procedure (e.g., the {PAGA} initialization or a fast-gradient based method applied to substitute $t$). These $T$ task-specific elites are then embedded into the initial population, complemented by additional randomly initialized individuals. As a result, the initial population contains both broad exploratory diversity and task-informed individuals for all $T$ tasks, which enhances subsequent cross-task knowledge transfer.

{
The $T$ substitute autoencoders used by PAMFGA are trained independently
for the same communication configuration $(\eta,k)$ and channel model.
They share the same overall end-to-end communication objective but are
initialized independently and trained as separate models, resulting in
different learned encoder--decoder parameters and decision boundaries.
The substitute models are not copies of the unknown target and no target
parameters or queries are used during their construction.
In the experiments, the substitute autoencoders follow the MLP-based
architecture described in Section~6.1.  Their training data are independently generated from the same uniform message
distribution and communication setting. Thus, the surrogate diversity used
during PAMFGA optimization primarily arises from independent model
initialization and training rather than from assuming knowledge of the target
architecture. 
}

\subsection{Multifactorial operators}
\label{subsec:PAMFGA_operators}

The evolutionary search mechanism in {PAMFGA} is grounded in the multifactorial optimization paradigm of MFEA-II~\cite{bali2019multifactorial}, which enables simultaneous optimization over multiple tasks by means of implicit genetic transfer. Each individual $\mathbf{ind}$ is linked to a skill factor $\tau \in \{1,2,\ldots,T\}$ denoting the task for which its fitness is evaluated.

A central component of the framework is the dynamic {RMP} matrix
\begin{equation}
    \mathsf{RMP} \in [0,1]^{T \times T},
\end{equation}
where each entry $\mathsf{RMP}_{k,k'}$ controls the probability of performing inter-task mating between tasks $k$ and $k'$. The matrix is symmetric, i.e., $\mathsf{RMP}_{k,k'} = \mathsf{RMP}_{k',k}$, and is initialized with relatively high values (e.g., $0.95$) to encourage aggressive cross-task exploration at the early stages of the search.

During the mating phase, a pair of parents $(\mathbf{ind}_a,\mathbf{ind}_b)$ with skill factors $\tau_a$ and $\tau_b$ is selected:
\begin{itemize}
    \item If $\tau_a = \tau_b$, an \emph{intra-task} crossover is always performed, and the offspring are evaluated on that shared task.
    \item If $\tau_a \neq \tau_b$, \emph{inter-task} crossover between tasks $\tau_a$ and $\tau_b$ is carried out with probability $\mathsf{RMP}_{\tau_a,\tau_b}$. Otherwise, mating is restricted to within-task reproduction.
\end{itemize}

After offspring evaluation, the corresponding {RMP} entry is updated adaptively to reflect the quality of inter-task knowledge transfer. Denote by $f_o$ and $f_p$ the fitness of an offspring and its parent on the inherited task, respectively. The {RMP} value is updated according to
\begin{equation}
    \mathsf{RMP}_{\tau_a, \tau_b}
    \leftarrow
    \begin{cases}
        \min\bigl(1.0,\ \mathsf{RMP}_{\tau_a,\tau_b} + \Delta_{\text{inc}}\bigr), & \text{if } f_o > f_p,\\[0.2em]
        \max\bigl(0.1,\ \mathsf{RMP}_{\tau_a,\tau_b} - \Delta_{\text{dec}}\bigr), & \text{otherwise},
    \end{cases}
\end{equation}
where $\Delta_{\text{inc}}, \Delta_{\text{dec}} > 0$ are small step sizes controlling the adaptation rate. In this way, task pairs that benefit from mutual knowledge transfer (i.e., produce improved offspring) gradually obtain higher {RMP} values, whereas unhelpful task pairings are discouraged.

\subsubsection*{Crossover mechanisms}

Within each mating event, {PAMFGA} primarily employs the real-coded crossover operators already introduced in Section~\ref{sec:white-box-algorithm} for {PAGA}, namely {HX} and {SBX}. These operators operate in the USS and are applied to the real-valued chromosomes to generate offspring that balance the exploitation of parental structures with the exploration of new perturbation patterns. After crossover, each offspring chromosome is normalized to satisfy the perturbation power constraint.

To further enrich diversity, {PAMFGA} incorporates an additional {dOX} operator. In dOX, a contiguous window of length
\begin{equation}
    w = W \cdot (2\eta), \quad W \in [0,1],
\end{equation}
is used to define the segment exchanged between parents, where the window ratio $W$ is modulated by the current {RMP} value of the corresponding task pair. High {RMP} values, indicating strong task-relatedness, result in larger crossover windows and thus more substantial knowledge transfer. When the exchanged segment is too small to induce sufficient variation, a local 2-opt style permutation is applied to the offspring chromosome to re-order a subset of genes and increase structural diversity.

\subsubsection*{Mutation and scalar fitness}

Mutation introduces additional stochastic variation into the population and mitigates premature convergence. In {PAMFGA}, mutation is applied to offspring with probability $P_{m} \in [0,1]$. The mutation operators are consistent with those used in {PAGA}, namely {CIM}, which reverses the order of genes around a randomly chosen pivot, and {PM}, which perturbs gene values according to a bounded polynomial distribution. After mutation, chromosomes are renormalized to ensure feasibility with respect to the power constraint.

Selection is governed by scalar fitness, as in standard multifactorial evolutionary algorithms. For each individual, the raw fitness is first evaluated on its associated task, and then transformed into a scalar fitness value that allows comparison across tasks. Individuals with higher scalar fitness are more likely to survive and reproduce, ensuring that good task-specific solutions are retained while still allowing beneficial cross-task genetic material to proliferate. This mechanism maintains a balance between exploitation within each task and exploration through inter-task knowledge transfer.

\subsection{Post-processing of elite perturbations}
\label{sec:svd}

After $G_{\max}$ generations of evolution, {PAMFGA} returns, for each of the $T$ tasks, a best (elite) perturbation vector. Denote this set of task-wise elite perturbations by
\begin{equation}
    \mathcal{P} = \{\mathbf{p}^{1}, \mathbf{p}^{2}, \ldots, \mathbf{p}^{T}\},
\end{equation}
where $\mathbf{p}^{j} \in \mathbb{C}^{\eta}$ is the elite perturbation associated with substitute autoencoder $j$. The goal is to aggregate these task-wise perturbations into a single universal perturbation suitable for attacking the unknown target autoencoder in a black-box fashion.

To this end, a matrix of normalized perturbations is first formed as
\begin{equation}
    \mathbf{P}^{\text{norm}}
    =
    \begin{bmatrix}
        \dfrac{\mathbf{p}^{1}}{\|\mathbf{p}^{1}\|_{2}} &
        \dfrac{\mathbf{p}^{2}}{\|\mathbf{p}^{2}\|_{2}} &
        \cdots &
        \dfrac{\mathbf{p}^{T}}{\|\mathbf{p}^{T}\|_{2}}
    \end{bmatrix}^{\top},
\end{equation}
where each row corresponds to a unit-norm perturbation direction in the complex signal space. Even randomly shifted versions of these elite perturbations typically induce substantial misclassification on their respective substitute models, suggesting that their dominant shared structure captures useful adversarial directions.

{
Let $\mathbf{q}_j=\mathbf{p}_j/\|\mathbf{p}_j\|_2$
denote the normalized elite perturbation associated with substitute autoencoder $j$. The normalized perturbations matrix $\mathbf{P}_{\mathrm{norm}}$ has singular value decomposition written as
\begin{equation}
\mathbf{P}_{\mathrm{norm}}
=
\mathbf{U}\boldsymbol{\Sigma}\mathbf{V}^{\mathrm H},
\end{equation}
where $(\cdot)^{\mathrm H}$ denotes the conjugate transpose. The first right
singular vector $\mathbf{v}_1$ solves
\begin{equation}
\mathbf{v}_1
=
\underset{\|\mathbf{v}\|_2=1}{\arg\max}
\sum_{j=1}^{T}
\left|\mathbf{q}_j^{\mathrm H}\mathbf{v}\right|^2.
\end{equation}
Hence, $\mathbf{v}_1$ represents the unit-norm direction with the maximum
aggregate squared alignment with the elite perturbations obtained from all
substitute models. It should therefore be interpreted as a dominant consensus direction rather than as a direction that is guaranteed to be
optimal for the unknown target autoencoder. }

{
The motivation is based on the transferability assumption that adversarial components shared by several independently trained substitute models are
more likely to generalize to an unknown target than components specific to
one substitute model. Compared with direct averaging, the squared-alignment
criterion is not affected by cancellation caused by sign or global-phase ambiguities among the elite perturbations. Compared with selecting the best
single-surrogate perturbation, it does not require choosing one substitute
model as representative of the unknown target. Moreover, SVD is applied only as a low-cost post-processing operation, whereas directly optimizing an ensemble objective would require evaluating each candidate on all substitute
models.
The proposed SVD aggregation is thus a computationally efficient, target-agnostic heuristic for extracting shared adversarial structure. It does not theoretically guarantee the strongest possible attack on every unknown target, nor do we claim that it is universally superior to all other aggregation strategies. After appropriate rescaling to meet the power constraint, this vector is transmitted by the adversary to attack the unknown target autoencoder.}

\begin{remark}[Two level transfer learning]
In this paper, the role of transfer learning is embodied in the second proposed framework, PAMFGA, and is realized at two distinct levels. At the first level, a multitask evolutionary framework is adopted, where a single population is maintained to solve \(T\) tasks simultaneously. When selecting parents for crossover, if the parents have different skill factors \(k\) and \(k'\), mating occurs with probability \(\mathrm{RMP}(k,k')\); this mechanism effectively enables knowledge transfer between individuals specialized in different tasks. At the second level, in the black-box attack scenario, the attacker has no access to the encoder or decoder of the target end-to-end communication system. The proposed strategy is that the attacker first trains surrogate models, applies the evolutionary algorithm to attack these surrogate models, and then uses an SVD-based decomposition to aggregate the obtained solutions and transfer them to attack the target model. Experimental results demonstrate that this two-level transfer learning mechanism substantially enhances the effectiveness of the physical adversarial attack.
\end{remark}

\begin{remark}[Sensitivity of PAMFGA to the number of surrogate models]
{
The number of substitute models $T$ also affects the SVD-based aggregation,
because it determines the number of task-wise elite perturbations used to
construct $\mathbf{P}_{\mathrm{norm}}$. With a small $T$, the estimated
consensus direction is supported by only a limited set of surrogate-specific
perturbations and may therefore be more sensitive to individual surrogate
models. Increasing $T$ provides additional perturbation directions from which
shared adversarial structure can be extracted. However, the benefit is not
expected to grow indefinitely, since additional surrogate models may contribute
redundant or weakly aligned directions, leading to diminishing returns.
This behavior is consistent with the empirical results in Section~6.3:
increasing $T$ from $4$ to $10$ generally improves the transferred attack,
whereas the difference between $T=10$ and $T=14$ is considerably smaller.
We emphasize that varying $T$ changes both the multifactorial optimization
process and the set of perturbations supplied to the SVD stage; therefore,
these results characterize the sensitivity of the overall PAMFGA pipeline to
the number of surrogate models rather than an isolated SVD-only ablation.
}\end{remark}

\subsection{Computational complexity and discussion}\label{sec:computational-complexity}

The computational complexity of the proposed algorithms is now summarized, using basic arithmetic operations (addition, subtraction, multiplication, comparison, and array indexing) as $\mathcal{O}(1)$ primitives.
For {PAGA}, let $\eta$ denote the number of complex channel uses per symbol (so that each chromosome has length $2\eta$ in the real domain), $\mathsf{Pop_{size}}$ the population size, $G$ the total number of generations, and $k$ the number of bits per symbol. The initialization phase generates $\mathsf{Pop_{size}}$ individuals, each requiring $\mathcal{O}(\eta)$ operations for sampling and normalization, leading to an initialization cost of
\begin{equation}
    \mathcal{O}(\mathsf{Pop_{size}}\,\eta).
\end{equation}

In each generation, crossover and mutation operators process each individual with cost $\mathcal{O}(\eta)$, resulting in $\mathcal{O}(\mathsf{Pop_{size}}\,\eta)$ operations per generation for variation. Fitness evaluation dominates the complexity: for each individual, the BLER approximation is computed by simulating all $M = 2^{k}$ possible symbols and passing them through the autoencoder and the channel. Denote by $\mathcal{C}_{N}$ the cost of one forward pass (including channel and decoder), then the fitness evaluation cost per generation is
\begin{equation}
    \mathcal{O}\bigl(\mathsf{Pop_{size}}\,2^{k}\,\mathcal{C}_{N}\bigr).
\end{equation}
Sorting the population for selection requires $\mathcal{O}(\mathsf{Pop_{size}}\log \mathsf{Pop_{size}})$ operations. Therefore, the total complexity of {PAGA} over $G$ generations is
\begin{equation}
    \mathcal{O}\!\left(
        G\,\mathsf{Pop_{size}}\bigl(
            \eta + 2^{k}\mathcal{C}_{N} + \log \mathsf{Pop_{size}}
        \bigr)
    \right).
\end{equation}

For {PAMFGA}, the per-generation complexity is of the same order, because the chromosome representation, variation operators, and fitness evaluation per individual are analogous. The main additional costs arise from (i) training the $T$ substitute autoencoder models and (ii) performing the {SVD} in the post-processing stage. Let $\mathcal{C}_{\text{train}}$ denote the training cost of a single substitute autoencoder. Training all $T$ models requires
\begin{equation}
    \mathcal{O}\bigl(T\,\mathcal{C}_{\text{train}}\bigr).
\end{equation}
The {SVD} on the normalized perturbation matrix $\mathbf{P}^{\text{norm}} \in \mathbb{C}^{T \times \eta}$ (or its real-valued counterpart of size $T \times 2\eta$) incurs a cost of
\begin{equation}
    \mathcal{O}\bigl(\eta T \min\{2\eta,T\}\bigr).
\end{equation}
Combining all terms, the overall complexity of {PAMFGA} is expressed as
\begin{equation}
    \mathcal{O}\!\left(
        T\,\mathcal{C}_{\text{train}}
        + G\,\mathsf{Pop_{size}}\bigl(
            \eta + 2^{k}\mathcal{C}_{N} + \log \mathsf{Pop_{size}}
        \bigr)
        + \eta T \min\{2\eta,T\}
    \right).
\end{equation}

\begin{remark}
The proposed {PAGA} and {PAMFGA} frameworks provide a unified and flexible evolutionary optimization approach to physical adversarial attack design in both white-box and black-box scenarios. In the white-box case, {PAGA} exploits full knowledge of the autoencoder architecture to directly optimize universal perturbations with high attack efficacy under power constraints. In the black-box case, {PAMFGA} relies on substitute models and multifactorial optimization to leverage transferability across tasks and learn robust perturbations without access to the target model. This comparison highlights that white-box evolutionary attacks benefit from detailed model awareness, whereas black-box evolutionary attacks emphasize adaptability and cross-task knowledge transfer, making them suitable for realistic deployment conditions. Together, the two algorithms offer a comprehensive toolset for adversarial design against DL–based communication systems.
\end{remark}

\section{Experimental Results}\label{sec:numerical_results}
\subsection{Experimental settings}

The numerical experiments are conducted over an additive white Gaussian noise (AWGN) channel. The channel noise $n(t)$ is modeled as an ergodic random process added to the transmitted signal, with constant two-sided power spectral density $G_n(f) = N_0/2$ and samples following a zero-mean complex Gaussian distribution. In the absence of adversarial perturbations, this model corresponds to the standard AWGN channel widely adopted in digital communication analysis.

The underlying communication system is implemented using an MLP-based autoencoder architecture \cite{goodfellow2016deep}, following the design principles in \cite{o2017introduction,sadeghi2019physical,son2024adversarial}. The autoencoder is configured for several $(\eta,k)$ pairs, namely
\begin{equation}(\eta,k) \in \{(5,3), (7,4), (11,6)\},
\end{equation} where $k$ denotes the number of information bits per symbol and $\eta$ the number of channel uses. For each configuration, the encoder maps messages from a constellation of size $M = 2^{k}$ to a $2\eta$-dimensional real-valued transmit vector that satisfies the average power constraint, while the decoder implements a multi-class classifier over the $M$ possible messages.

To represent the energy constraint in \eqref{eq:bler_max_attack}, the perturbation power is parameterized via the {PSR}, defined as the ratio between the perturbation energy and the signal energy, expressed in decibels. In the experiments, three PSR levels are considered: $\mathrm{PSR} \in \{-2,-6,-10\}$~dB, corresponding to relatively strong, moderate, and weak perturbations, respectively. During autoencoder training and during the fitness approximation of the attack methods, the {SNR} is set to a fixed value of $1$~dB, while performance curves in the subsequent sections are reported over a range of {SNR} values.

For comparative analysis, the proposed evolutionary attack algorithms are evaluated against several existing physical adversarial attack strategies. Specifically, the following baselines are considered:
\begin{itemize}
    \item FGMA, the fast-gradient based physical attack introduced in \cite{sadeghi2019physical};
    \item RSBA, the randomly shifted black-box attack from \cite{sadeghi2019physical};
    \item RMAEP, the projected-gradient–based physical attack for RIS-assisted MIMO autoencoders proposed in \cite{son2024adversarial};
    \item a Gaussian jamming attack, where the perturbation is drawn as a complex Gaussian noise vector with power matched to the chosen PSR.
\end{itemize}
{FGMA and RMAEP are primarily designed for white-box settings. For the black-box comparison, RMAEP is adapted by applying it independently to the substitute autoencoders and then aggregating the resulting perturbations using the SVD-based procedure described in Section~5.3. Therefore, the RMAEP
results reported in the black-box experiments correspond to this SVD-adapted variant rather than to the original white-box RMAEP. The same SVD
aggregation rule is employed to provide a common surrogate-to-target transfer protocol for the compared perturbation-generation methods. For brevity and to avoid introducing additional abbreviations, we retain the label ``RMAEP'' in the black-box figures and tables; throughout these results, this label refers to the above SVD-adapted variant.}

All simulations are implemented in Python and executed on a machine equipped with an Intel Core i5 v4 processor running at 2.20~GHz, 16~GB of RAM, and the Windows~11 operating system. Input data for the MLP-autoencoder are generated i.i.d. according to a uniform distribution over the $M$ messages, yielding training datasets of size $(N,M)$. Each autoencoder model is trained for $1000$ epochs using mini-batches of size $1000$. The Adam optimizer is employed with an initial learning rate of $3 \times 10^{-4}$, and a cosine decay schedule is used to gradually reduce the learning rate during training.

For the proposed evolutionary attack algorithms, the hyperparameters are selected based on preliminary tuning over a validation set. Unless otherwise stated, both PAGA and PAMFGA use a population size of $\mathsf{Pop_{size}} = 100$, a maximum number of generations $G_{\max} = 250$, a crossover probability $p_c = 0.8$, a mutation probability $p_m = 0.2$, and an elite rate $p_e = 5\%$ of the population. The distribution index of the SBX operator is set to $\eta_c = 2$. In PAMFGA, the random mating probability matrix is initialized so that all off-diagonal entries take the value $0.7$, encouraging inter-task genetic exchange at the beginning of the search. These settings provide a good balance between exploration and exploitation and are used consistently across the experimental scenarios reported following.
\begin{figure}[pos=!b]
    \centering
    \begin{subfigure}{0.48\textwidth}
        \includegraphics[width=\textwidth]{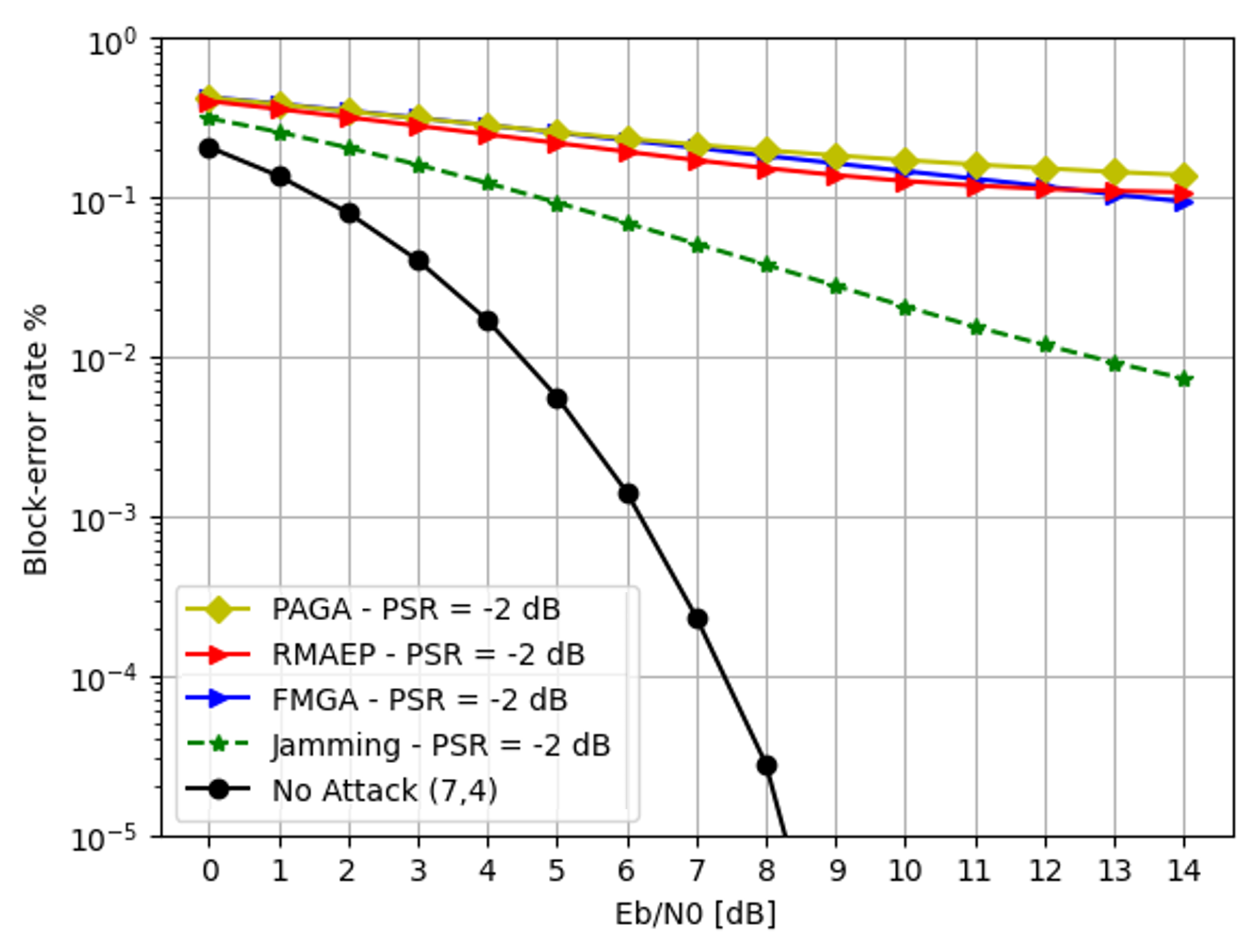}
        \caption{}
        \label{fig:result1}
    \end{subfigure}
    \hfill
    \begin{subfigure}{0.48\textwidth}
        \includegraphics[width=\textwidth]{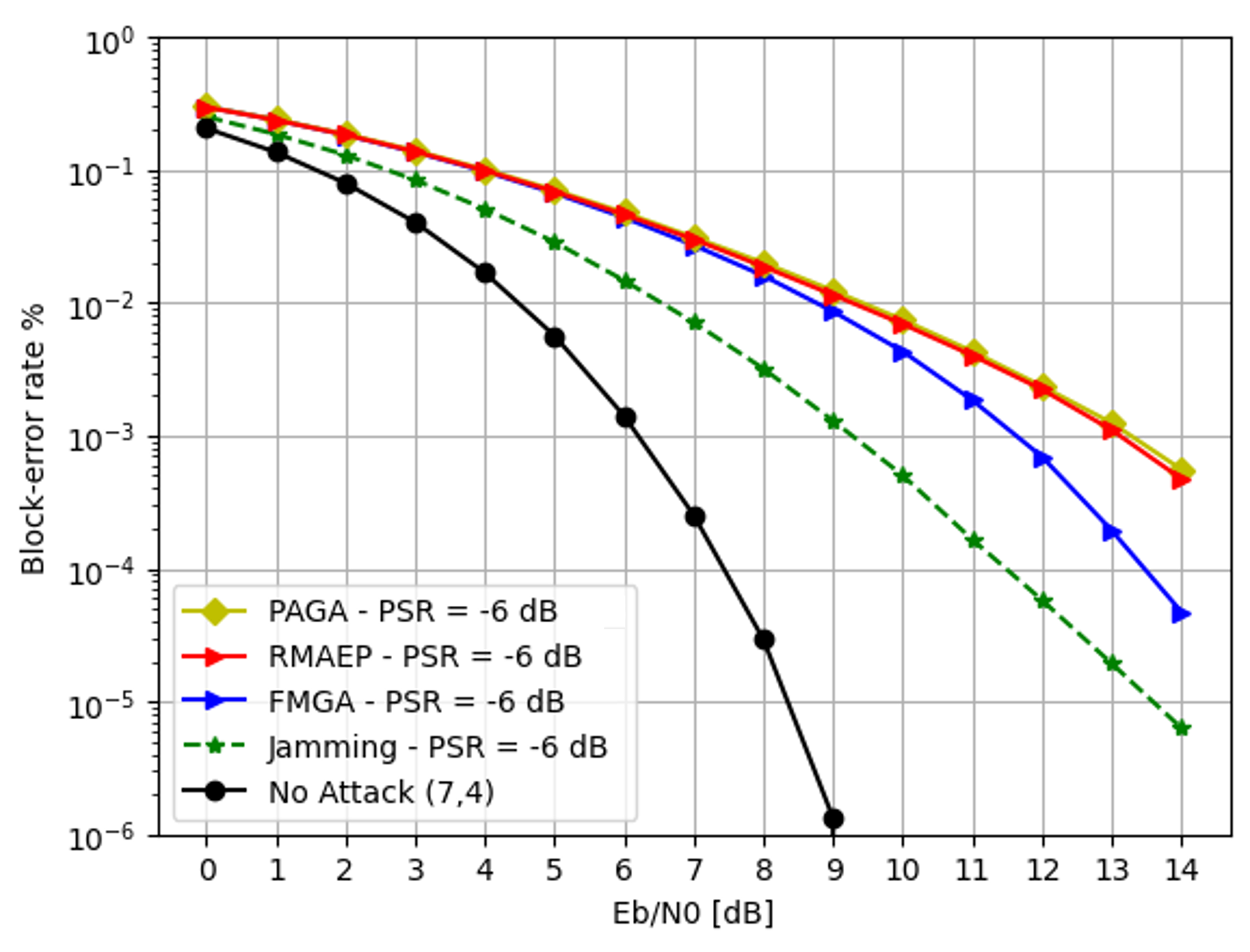}
       \caption{}
        \label{fig:result2}
    \end{subfigure}
    \hfill
    \begin{subfigure}{0.48\textwidth}
        \includegraphics[width=\textwidth]{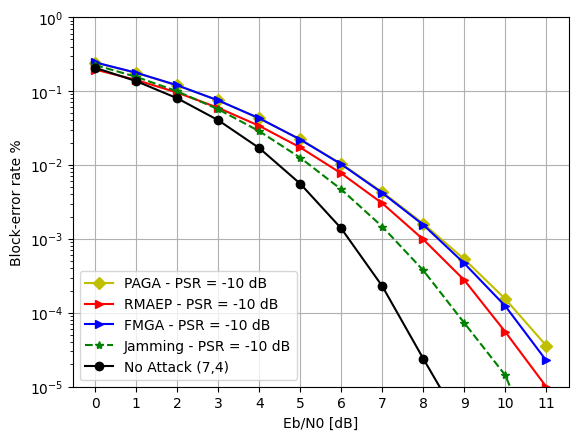}
       \caption{}
        \label{fig:result3}
    \end{subfigure}
    \begin{subfigure}{0.48\textwidth}
        \includegraphics[width=\textwidth]{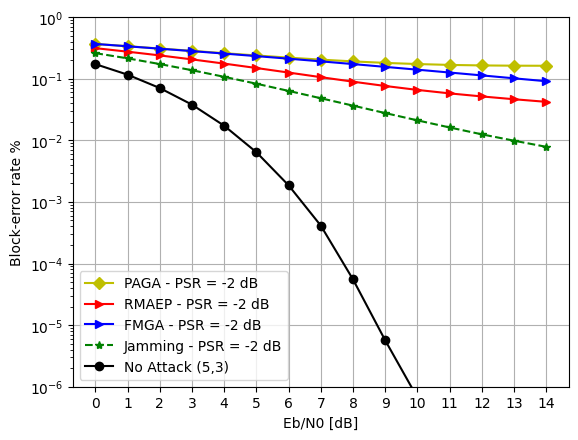}
        \caption{}
        \label{fig:result4}
    \end{subfigure}
 \label{fig:three_images}
\end{figure}
\begin{figure}[pos=!t]\ContinuedFloat
  \setcounter{subfigure}{4}
    \centering
     \begin{subfigure}{0.48\textwidth}
        \includegraphics[width=\textwidth]{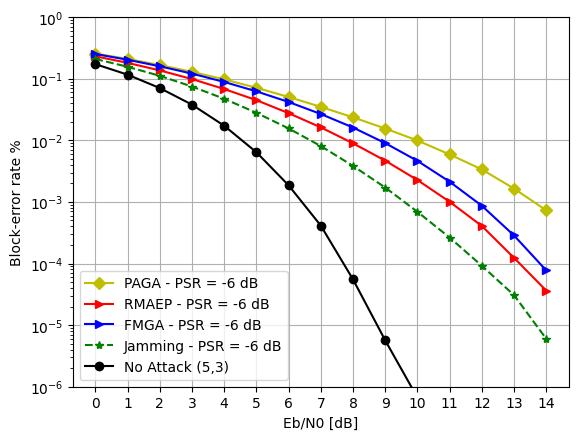}
        \caption{}
        \label{fig:result5}
    \end{subfigure}
     \begin{subfigure}{0.48\textwidth}
        \includegraphics[width=\textwidth]{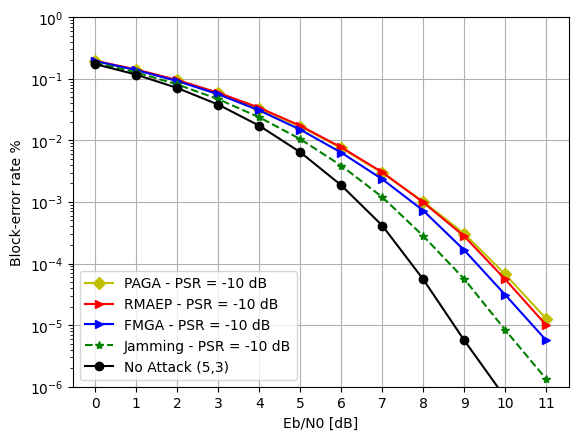}
        \caption{}
        \label{fig:result6}
    \end{subfigure}
    \begin{subfigure}{0.48\textwidth}
        \includegraphics[width=\textwidth]{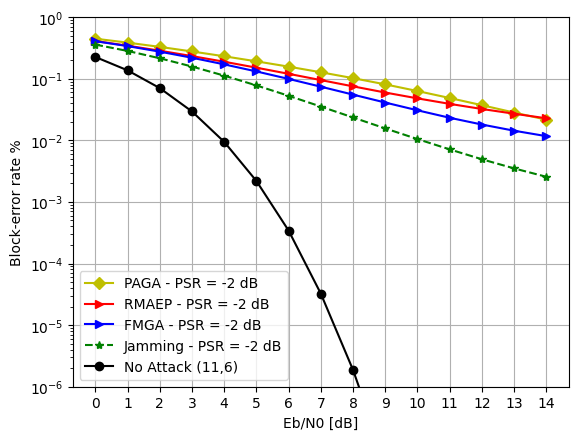}
        \caption{}
        \label{fig:result7}
    \end{subfigure}
     \begin{subfigure}{0.48\textwidth}
        \includegraphics[width=\textwidth]{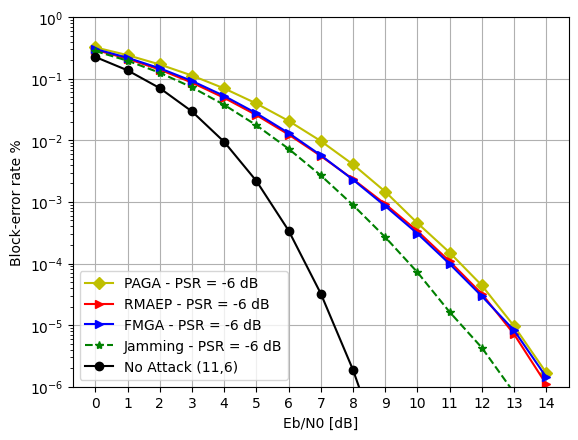}
        \caption{}
        \label{fig:result8}
    \end{subfigure}
     \begin{subfigure}{0.48\textwidth}
        \includegraphics[width=\textwidth]{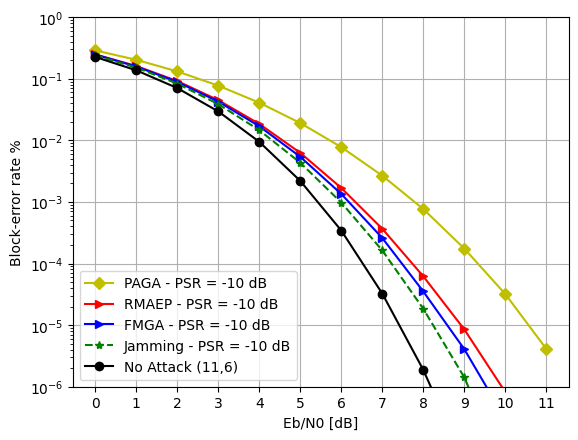}
        \caption{}
        \label{fig:result9}
    \end{subfigure}  
    \caption{BLER versus Eb/N0 against the MLP autoencoder by the algorithms across the varying systems and constraints.}
\label{fig:three_images-2}
\end{figure}
\subsection{Results analysis for white-box scenario}
This section first examines the performance of the proposed PAGA scheme in the white-box setting. Across all constellations and perturbation-to-signal ratios (PSRs) in Figs.~\ref{fig:result1}–\ref{fig:result9}, PAGA consistently yields the highest BLER, confirming its effectiveness as a model-aware physical-layer adversarial attack. Its impact is particularly striking when compared with conventional Gaussian jamming and the learning-based baselines FGMA and RMAEP.

For the strongest attack regime, $\mathrm{PSR} = -2$~dB in Figs.~\ref{fig:result1}, \ref{fig:result4}, and \ref{fig:result7}, PAGA causes a dramatic degradation of the autoencoder. For example, in Fig.~\ref{fig:result1} with $(\eta,k)=(7,4)$, PAGA attains a BLER on the order of $2\times 10^{-1}$ at moderate $\mathrm{E_b/N_0}$, whereas jamming only reaches about $9\times 10^{-3}$. Similar gaps are observed for $(\eta,k)=(5,3)$ and $(11,6)$, where PAGA outperforms FGMA and RMAEP by factors between roughly $2\times$ and $6\times$, depending on the SNR and parameter setting.
When the perturbation power is reduced to $\mathrm{PSR}=-6$~dB (Figs.~\ref{fig:result2}, \ref{fig:result5}, and \ref{fig:result8}), PAGA remains the most damaging strategy. In Fig.~\ref{fig:result5} with $(\eta,k)=(5,3)$, PAGA still induces a BLER of about $10^{-3}$, while jamming yields only around $10^{-5}$, i.e., more than two orders of magnitude lower. This demonstrates that PAGA retains a clear advantage even when the available perturbation energy is significantly constrained. Moreover, its superiority persists across different code parameters, indicating strong generalization over the considered autoencoder configurations.
At the most stringent power budget, $\mathrm{PSR}=-10$~dB (Figs.~\ref{fig:result3}, \ref{fig:result6}, and \ref{fig:result9}), all attacks naturally suffer a reduction in absolute BLER. Nevertheless, PAGA continues to dominate the competing methods. For instance, in Fig.~\ref{fig:result9} with $(\eta,k)=(11,6)$, PAGA achieves a BLER that is roughly an order of magnitude higher than RMAEP over a wide SNR range, despite the severe perturbation constraint. Even in this low-power regime, PAGA consistently maintains a logarithmic advantage over jamming, FGMA, and RMAEP.

\begin{figure}[pos=h]
    \centering
    \begin{subfigure}{0.48\textwidth}
        \includegraphics[width=\textwidth]{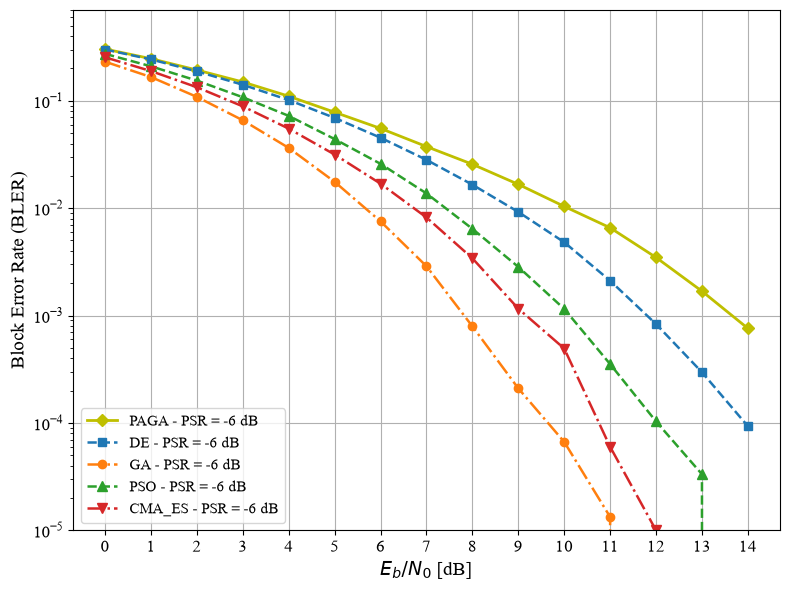}
        \caption{BLER versus $E_b/N_0$ under the matched evaluation budget.}
        \label{fig:white-heuristic}
    \end{subfigure} 
    \hfill
    \begin{subfigure}{0.48\textwidth}
        \includegraphics[width=\textwidth]{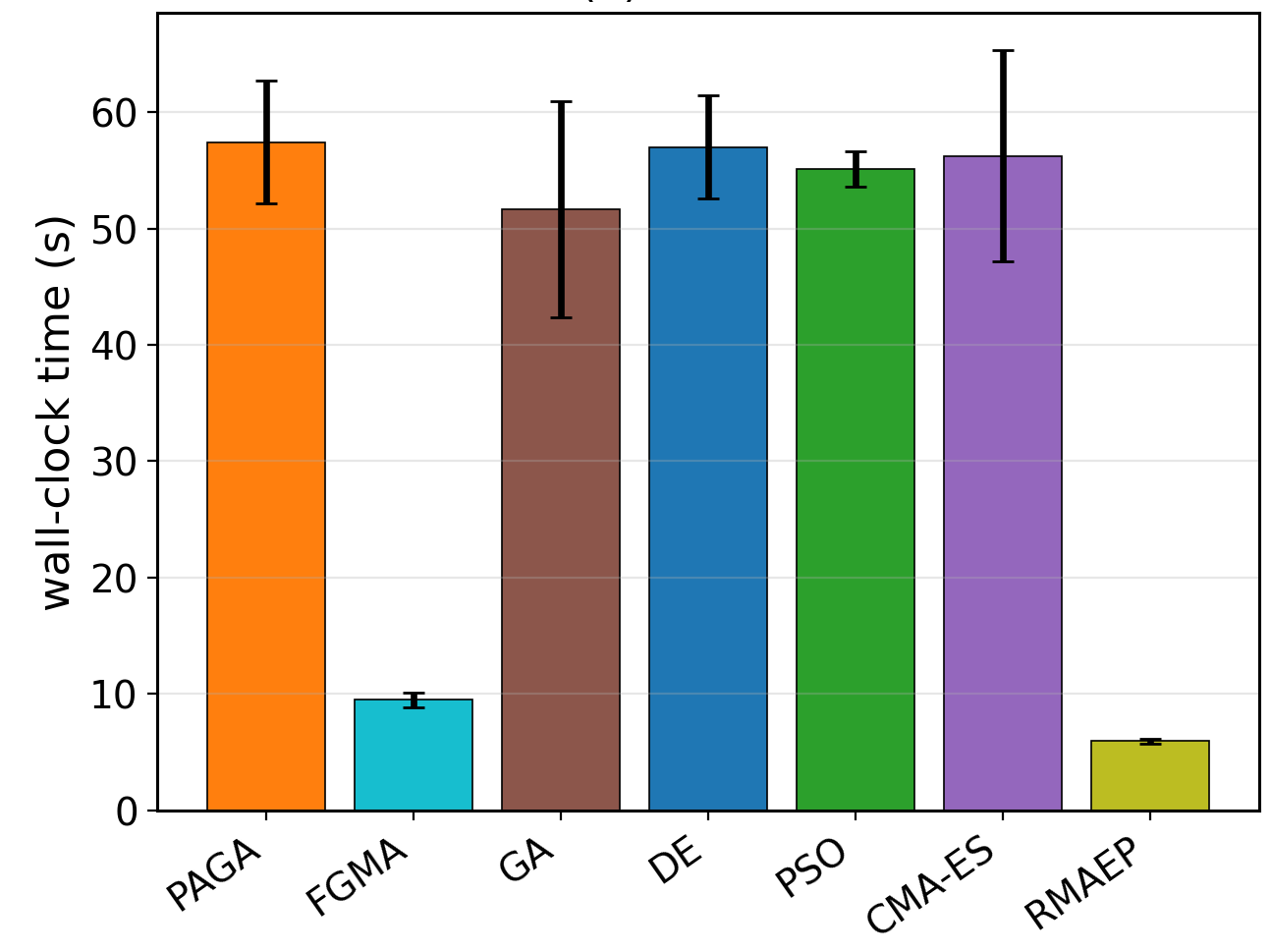}
        \caption{Wall-clock runtime in the white-box setting.}
        \label{fig:resources_whitebox}
    \end{subfigure}
    \caption{{White-box comparison of PAGA with representative
    attack and metaheuristic baselines at $\mathrm{PSR}=-6$ dB:
    (a) attack effectiveness under a matched fitness-evaluation budget for
    the population-based solvers; and (b) measured wall-clock runtime.
    Error bars in (b) indicate the variability across repeated runs.}}
    \label{fig:three_images}
\end{figure}

{
To further examine whether the advantage of PAGA results from the choice of
evolutionary solver or simply from a larger computational budget, we compare
it with representative population-based optimizers under the same
fitness-evaluation budget. As shown in Fig.~\ref{fig:white-heuristic},
PAGA generally maintains the highest BLER over the considered $E_b/N_0$
range at $\mathrm{PSR}=-6$ dB. The advantage becomes particularly visible
at medium and high SNRs, where the BLER obtained by the competing GA \cite{dang2025jamming}, DE \cite{lu2024energy},
PSO \cite{binh2026hybrid}, and CMA-ES \cite{saranya2025tdoa} solvers decreases more rapidly. This result indicates that
the attack effectiveness of PAGA cannot be attributed solely to using a
population-based search or to performing more fitness evaluations; rather,
its problem-specific initialization and evolutionary operators provide a
more effective search of the power-constrained perturbation space under the
matched evaluation budget.
Figure~\ref{fig:resources_whitebox} further reports the measured wall-clock
runtime in the same white-box setting. PAGA requires a runtime comparable
to those of the considered population-based metaheuristics, including DE,
PSO, and CMA-ES, and only moderately higher than the simple GA baseline.
In contrast, the gradient-based FGMA and RMAEP baselines are substantially
faster, reflecting their different optimization procedures. Therefore, we do
not claim that PAGA is the fastest attack method. Rather, the results show
that its stronger attack effectiveness relative to the other metaheuristic
solvers is achieved without a disproportionate increase in wall-clock cost.
Together, the matched-budget and runtime comparisons provide a more complete
view of the solution-quality--computation trade-off of PAGA in the white-box
scenario.}
Overall, these results confirm that PAGA is not only highly effective under generous PSR budgets but also remarkably robust under stringent power limitations. Across all studied settings, it yields BLERs that are frequently one to two orders of magnitude higher than those of conventional jamming and prior learning-based attacks, highlighting the serious security risk posed by tailored, model-aware adversarial perturbations in end-to-end autoencoder communication systems.

\paragraph{Sensitivity to stochastic fitness evaluation}
The fitness evaluation in Algorithm~1 is stochastic because the received
signal depends on the randomly sampled channel noise. Although using one noise
realization per transmitted message provides a low-cost estimate during
evolutionary optimization, a noisy fitness value may favor candidates that
happen to perform well for particular noise draws. We therefore conduct an
additional sensitivity analysis to examine whether this effect materially
changes the quality of the final perturbation. 
Let $N_{\mathrm{fit}}$ denote the number of independently sampled
channel-noise realizations used per transmitted message during a fitness
evaluation. For this analysis, the stochastic fitness estimator can be written
as
}
\begin{equation}
\widehat{\mathrm{BLER}}_{N_{\mathrm{fit}}}(\mathbf{p})
=
\frac{1}{M N_{\mathrm{fit}}}
\sum_{s\in\mathcal{M}}
\sum_{r=1}^{N_{\mathrm{fit}}}
\mathbf{1}
\left\{
g\!\left(
f(s)+\mathbf{n}_{s,r}+\mathbf{p}
\right)
\neq s
\right\},
\label{eq:fitness_sensitivity}
\end{equation}
where $\mathbf{n}_{s,r}$ denotes an independently generated channel-noise
realization. The default PAGA implementation corresponds to
$N_{\mathrm{fit}}=1$.

We compare $N_{\mathrm{fit}}\in\{1,5,10\}$ using the representative
configuration $(\eta,k)=(7,4)$, $\mathrm{PSR}=-6$ dB, and an optimization
SNR of $1$ dB. All remaining PAGA hyperparameters and the autoencoder
checkpoint are kept fixed. For each value of $N_{\mathrm{fit}}$, we perform
10 independent complete optimization runs. After each run, the final elite
perturbation is re-evaluated using
$N_{\mathrm{eval}}=1000$ fresh channel-noise realizations per transmitted
message that are generated independently of those used during optimization.
Since $M=2^4=16$, each final elite is therefore evaluated using 16,000 fresh
received samples. 
\begin{table}[]
    \centering
    \caption{
Sensitivity of PAGA to the number of noise realizations used during fitness
evaluation. Fresh-sample BLER is evaluated using
$N_{\mathrm{eval}}=1000$ unseen noise realizations per transmitted message.
Reported BLER intervals are 95\% confidence intervals (CI) over 10 independent runs.
}
\label{tab:Nfit}
    \setlength{\tabcolsep}{5pt}
\begin{tabular}{ccccc}
\toprule
$N_{\mathrm{fit}}$
& In-search fitness
& Fresh BLER (95\% CI)
& Selection gap
& Runtime (s) \\
\midrule
1
& 0.6813
& 0.2417 $[0.2312,\,0.2522]$
& 0.4396
& 11.14 \\
5
& 0.4475
& 0.2407 $[0.2345,\,0.2468]$
& 0.2068
& 25.67 \\
10
& 0.3875
& 0.2458 $[0.2384,\,0.2531]$
& 0.1417
& 68.55 \\
\bottomrule
\end{tabular}
\end{table}

The results in Table \ref{tab:Nfit} reveal two complementary effects. First, increasing
$N_{\mathrm{fit}}$ reduces the optimism of the stochastic fitness value used
during evolutionary selection. The mean difference between the selected
elite's in-search fitness and its subsequent fresh-sample BLER decreases from
$0.4396$ at $N_{\mathrm{fit}}=1$ to $0.2068$ and $0.1417$ at
$N_{\mathrm{fit}}=5$ and $10$, respectively. Thus, averaging over more
channel-noise realizations provides a less optimistic estimate of the
candidate's expected attack performance. 
Second, and more importantly, this reduction in selection optimism does not
produce a systematic improvement in the final perturbation when it is
evaluated on unseen noise samples. The mean fresh-sample BLER is $0.2417$,
$0.2407$, and $0.2458$ for $N_{\mathrm{fit}}=1$, $5$, and $10$,
respectively. Using paired runs and common fresh-evaluation noise samples, the
mean difference between $N_{\mathrm{fit}}=5$ and
$N_{\mathrm{fit}}=1$ is $-0.0010$, with a 95\% confidence interval of
$[-0.0138,0.0118]$. For $N_{\mathrm{fit}}=10$ relative to
$N_{\mathrm{fit}}=1$, the corresponding difference is $0.0041$, with a
95\% confidence interval of $[-0.0101,0.0183]$. Both intervals include zero,
so the experiment does not indicate a systematic improvement in fresh-sample
attack performance from using additional noise realizations during
optimization. 
The increased averaging also incurs additional computational cost: the mean
optimization runtime rises from $11.14$ s for $N_{\mathrm{fit}}=1$ to
$25.67$ s for $N_{\mathrm{fit}}=5$ and $68.55$ s for
$N_{\mathrm{fit}}=10$. Taken together, these results support retaining
$N_{\mathrm{fit}}=1$ as the default low-cost stochastic estimator in PAGA.
While this setting produces a more optimistic in-search fitness estimate, the
final perturbations generalize comparably to fresh channel-noise samples in
the considered sensitivity experiment. 
We conduct this diagnostic analysis on PAGA because it directly isolates the
stochastic fitness estimator in Algorithm~1. Repeating the same analysis within PAMFGA would additionally mix the effect of noise averaging with inter-task evolutionary transfer and SVD-based surrogate aggregation. 

\subsection{Black-box scenario comparisons}
 
\paragraph{Cross-architecture transferability}
{Before evaluating the multifactorial attack, we first examine whether an
adversarial perturbation constructed on one autoencoder architecture remains
effective against receivers with different network structures. We craft the
attack using \texttt{MLP\_base} as the source model and directly transfer the
same perturbation to four unseen target architectures: \texttt{MLP\_relu},
\texttt{MLP\_deep}, \texttt{MLP\_widedec}, and a CNN-based autoencoder.
The first three targets belong to the same broad MLP family but differ in
activation, depth, or decoder width, whereas the CNN constitutes an
out-of-family target. No target-specific optimization or adaptation is
performed.}

\begin{figure}[pos=!t]
    \centering
    \includegraphics[width=.8\linewidth]{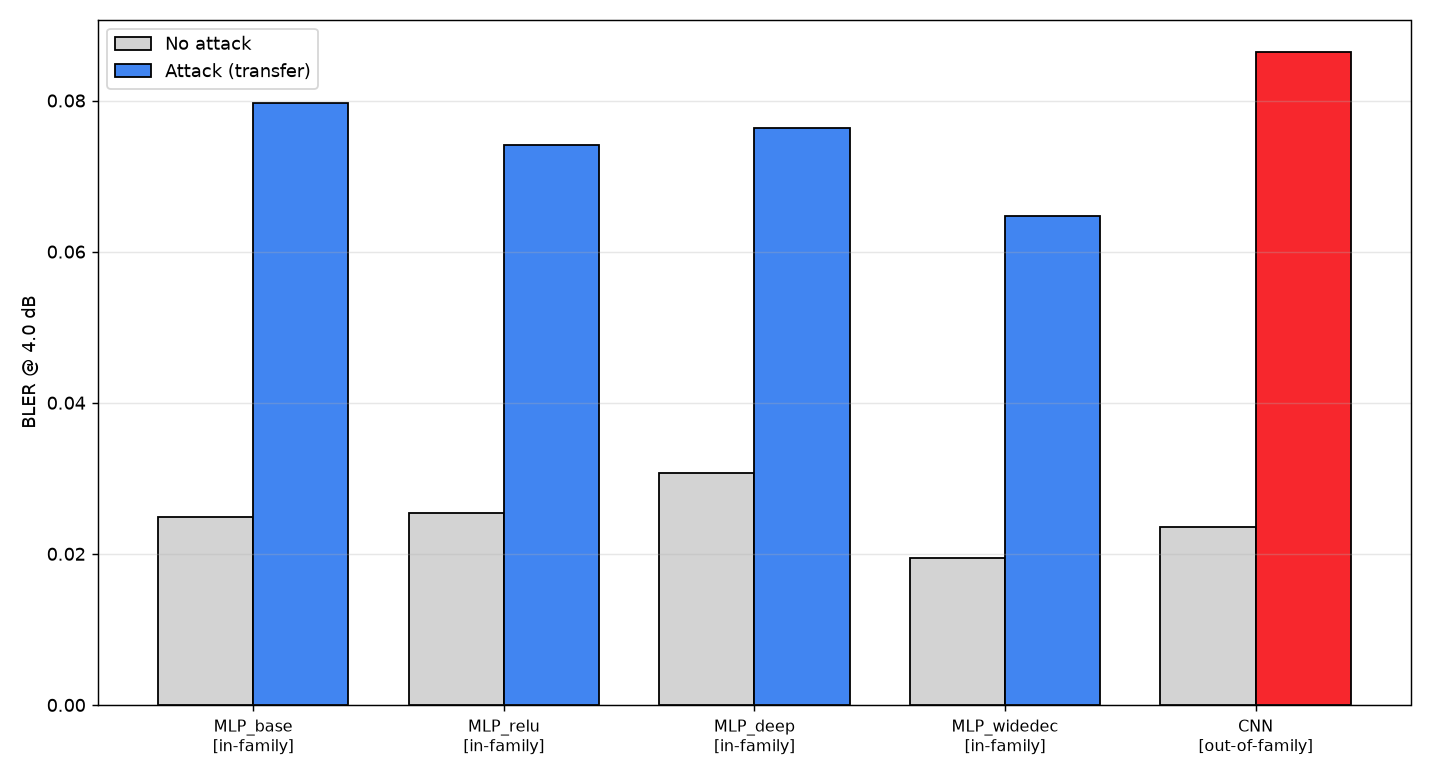}
    \caption {Cross-architecture attack transfer}
    \label{fig:cross_architecture}
\end{figure}

{As shown in Fig.~\ref{fig:cross_architecture}, the transferred perturbation
substantially increases the BLER for every target architecture at
$E_b/N_0=4$ dB and PSR $=-6$ dB. In particular, the BLER increases from
approximately 0.025 to 0.074--0.080 for the MLP variants and from
approximately 0.024 to 0.087 for the unseen CNN target. Interestingly, the
out-of-family CNN exhibits an attack-induced degradation comparable to, and
even larger than, those observed for the in-family targets. These results
indicate that the learned adversarial direction is not merely exploiting
architecture-specific characteristics of the source MLP, but captures
features that remain effective across substantially different receiver
architectures. This observation empirically supports the cross-model
transferability assumption underlying our limited-knowledge black-box
attack.}

\paragraph{PAMFGA solution quality and convergence}
Having established the cross-architecture transferability of adversarial
perturbations, we next investigate how effectively PAMFGA exploits multiple
surrogate models through multifactorial optimization. In particular, we
examine the impact of the number of simultaneously solved tasks, $T$.
Table~\ref{tab:changingT} reports both BLER and running time for different
values of $T$, and compares PAMFGA with a baseline strategy in which PAGA is
executed $T$ times independently and the resulting perturbations are combined
via SVD. Across the considered configurations, PAMFGA generally attains the highest
or a competitive BLER, indicating that joint optimization can provide
benefits over independently solving the surrogate tasks and aggregating
their solutions a posteriori.

{
A closer look at Table~\ref{tab:black-results} shows that increasing the
number of surrogate tasks from $T=4$ to $T=10$ is associated with higher BLER
over most of the considered $E_b/N_0$ values. However, the difference between
$T=10$ and $T=14$ is much smaller, suggesting diminishing returns when
additional surrogate models are introduced. This result indicates that, under
the considered experimental setting, the benefit of enlarging the surrogate
set tends to saturate beyond approximately $T=10$. Therefore, a moderate
number of surrogate models can provide a favorable empirical trade-off between
attack effectiveness and computational cost, although it should not be claimed that
$T=10$ is globally optimal.
}

\begin{table*}[t]
\centering
\caption{Performance of the compared algorithms across the considered system
configurations in the black-box attack scenario.}\label{tab:black-results}
\resizebox{\linewidth}{!}{
\begin{tabular}{cccccccccc}
\hline
                        & $E_b/N_0$      & 0        & 1        & 2        & 3        & 4        & 5        & 6        & 7        \\ \hline
\multirow{3}{*}{$T=14$} 
  & \textbf{PAMFGA}     & \textbf{0.3364} & \textbf{0.2898} & \textbf{0.2216} & \textbf{0.1684} &
                         \textbf{0.1245} & \textbf{0.0967} & \textbf{0.0750} & \textbf{0.0579} \\
  & RSBA                & 0.3072 & 0.2509 & 0.2007 & 0.1585 & 0.1242 & 0.0964 & 0.0748 & 0.0579 \\
  & PAGA                & 0.3332 & 0.2703 & 0.2132 & 0.1652 & 0.1245 & 0.0917 & 0.0719 & 0.0565 \\ \hline
\multirow{3}{*}{$T=10$} 
  & \textbf{PAMFGA}     & \textbf{0.3356} & \textbf{0.2892} & \textbf{0.2211} & \textbf{0.1682} &
                         \textbf{0.1243} & \textbf{0.0963} & \textbf{0.0747} & \textbf{0.0578} \\
  & RSBA                & 0.3064 & 0.2498 & 0.1993 & 0.1581 & 0.1242 & 0.0960 & 0.0745 & 0.0575 \\
  & PAGA                & 0.3329 & 0.2701 & 0.2129 & 0.1648 & 0.1240 & 0.0914 & 0.0716 & 0.0564 \\ \hline
\multirow{3}{*}{$T=4$}  
  & \textbf{PAMFGA}     & \textbf{0.2876} & \textbf{0.2283} & \textbf{0.1762} & \textbf{0.1366} &
                         \textbf{0.0963} & \textbf{0.0681} & \textbf{0.0475} & \textbf{0.0322} \\
  & RSBA                & 0.2867 & 0.2279 & 0.1760 & 0.1316 & 0.0958 & 0.0680 & 0.0470 & 0.0311 \\
  & PAGA                & 0.2872 & 0.2279 & 0.1762 & 0.1345 & 0.0961 & 0.0680 & 0.0473 & 0.0318 \\ \hline
\end{tabular}}

\vspace{0.3cm}
\resizebox{\linewidth}{!}{
\begin{tabular}{cccccccccc}
\hline
                        & $E_b/N_0$      & 8        & 9        & 10       & 11       & 12       & 13       & 14    & 15   \\ \hline
\multirow{3}{*}{$T=14$} 
  & \textbf{PAMFGA}     & \textbf{0.0452} & \textbf{0.0357} & \textbf{0.0290} & \textbf{0.0238} &
                         \textbf{0.0197} & \textbf{0.0164} & \textbf{0.0140} & \textbf{0.0119}\\
  & RSBA                & 0.0445 & 0.0342 & 0.0261 & 0.0197 & 0.0148 & 0.0108 & 0.0077 & 0.0058\\
  & PAGA                & 0.0430 & 0.0338 & 0.0274 & 0.0200 & 0.0152 & 0.0113 & 0.0087 & 0.0064\\ \hline
\multirow{3}{*}{$T=10$} 
  & \textbf{PAMFGA}     & \textbf{0.0450} & \textbf{0.0352} & \textbf{0.0287} & \textbf{0.0234} &
                         \textbf{0.0194} & \textbf{0.0162} & \textbf{0.0140} & \textbf{0.0118}\\
  & RSBA                & 0.0441 & 0.0339 & 0.0259 & 0.0196 & 0.0142 & 0.0105 & 0.0077 & 0.0057\\
  & PAGA                & 0.0440 & 0.0337 & 0.0274 & 0.0196 & 0.0150 & 0.0109 & 0.0086 & 0.0064\\ \hline
\multirow{3}{*}{$T=4$}  
  & \textbf{PAMFGA}     & \textbf{0.0216} & \textbf{0.0125} & \textbf{0.0075} & \textbf{0.0046} &
                         \textbf{0.0039} & \textbf{0.0013} & \textbf{0.0005} & \textbf{0.0004}\\
  & RSBA                & 0.0200 & 0.0123 & 0.0071 & 0.0040 & 0.0022 & 0.0010 & 0.0004 & 0.0003\\
  & PAGA                & 0.0209 & 0.0123 & 0.0074 & 0.0043 & 0.0027 & 0.0012 & \textbf{0.0005} & 0.0003\\ \hline
\end{tabular}
}

\end{table*}

\begin{figure}[pos=ht!]
    \centering
    \begin{subfigure}{0.48\textwidth}
        \includegraphics[width=\textwidth]{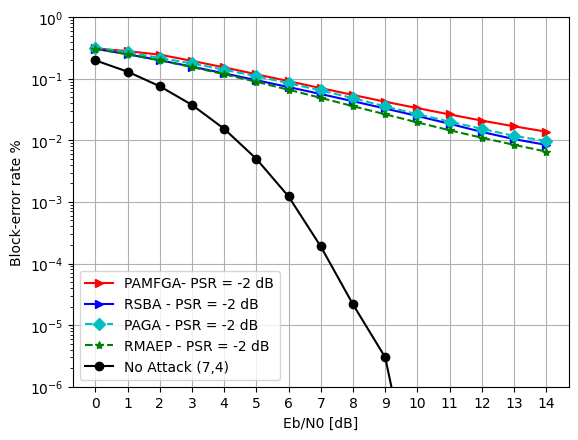}
       
        \label{fig:image1}
    \end{subfigure}
    \hfill
    \begin{subfigure}{0.48\textwidth}
        \includegraphics[width=\textwidth]{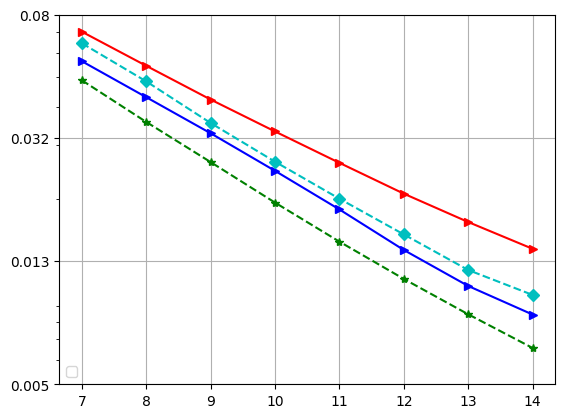}
    
        \label{fig:image2_7}
    \end{subfigure}
    \hfill
    \begin{subfigure}{0.48\textwidth}
        \includegraphics[width=\textwidth]{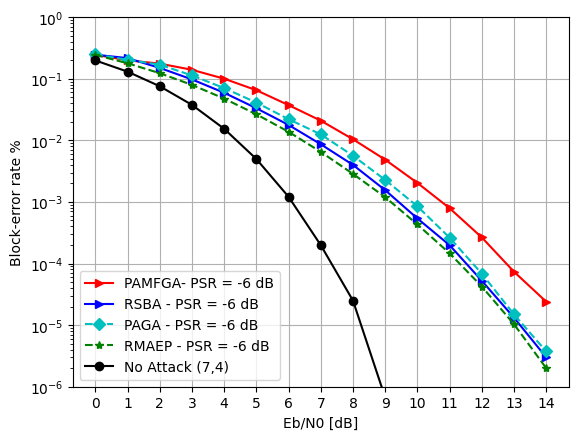}
        
        \label{fig:6}
    \end{subfigure}
    \hfill
    \begin{subfigure}{0.48\textwidth}
        \includegraphics[width=\textwidth]{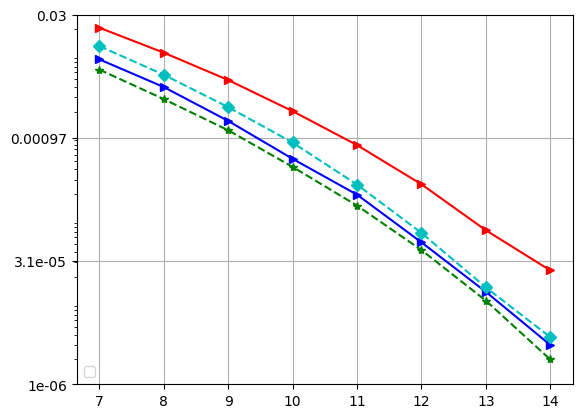}
        
        \label{fig:image6_7}
    \end{subfigure}
    \hfill
    \begin{subfigure}{0.48\textwidth}
        \includegraphics[width=\textwidth]{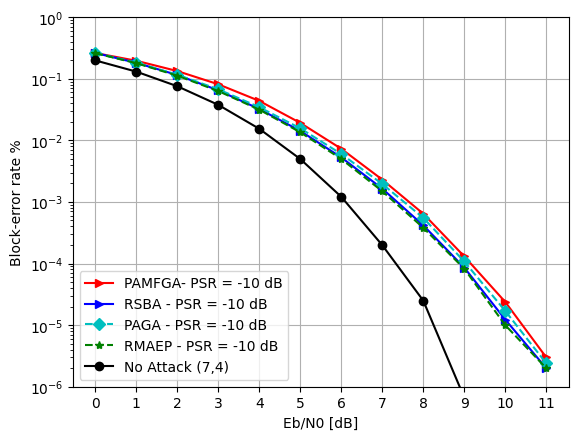}
    
        \label{fig:image10}
    \end{subfigure}
    \hfill
    \begin{subfigure}{0.48\textwidth}
        \includegraphics[width=\textwidth]{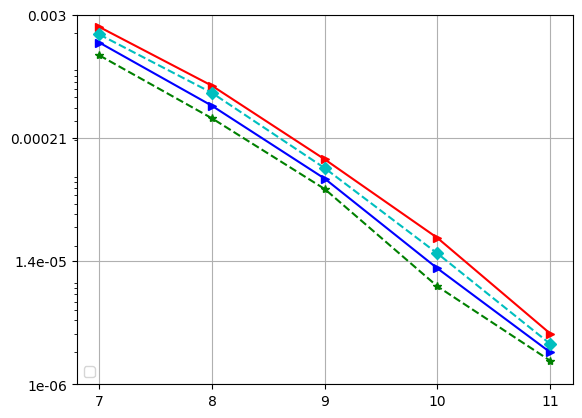}
    
        \label{fig:image10_7}
    \end{subfigure}
    \caption{BLER versus Eb/N0 in the black-box attacks against the CNN autoencoder with $T = 14$.}
    \label{fig:PAGA-BLER}

\end{figure}

Figure~\ref{fig:PAGA-BLER} illustrates BLER versus $E_b/N_0$ for the
CNN-based autoencoder under black-box attacks with $T = 14$.
{
PAMFGA generally achieves the highest or a competitive BLER over the
considered SNR range, although the numerical differences from RSBA
\cite{sadeghi2019physical}, RMAEP~\cite{son2024adversarial}, and the
SVD-based aggregation of PAGA are small at several operating points
}
\footnote{{In the black-box experiments, ``RMAEP'' and
``PAGA'' denote surrogate-based adapted variants rather than their original
white-box forms. For both baselines, perturbations are generated independently
on the substitute autoencoders and subsequently aggregated using the SVD
procedure in Section~\ref{sec:svd}. The short labels are retained to avoid
introducing additional abbreviations in the figures and tables.}}.
{
It should also be noted that the BLER axis is displayed on a logarithmic
scale to visualize attack performance over several orders of magnitude.
Therefore, the visual separation between the curves should not be interpreted
directly as a large absolute performance gap. At medium and high SNRs,
PAMFGA nevertheless maintains a generally higher BLER than the competing
methods, indicating that the transferred perturbation remains effective in
operating regions where the clean communication system would otherwise be
highly reliable.
}
{
These results provide evidence that the multitask evolutionary design can
discover perturbations that transfer from the surrogate models to the unseen
CNN decoder, while the magnitude of the improvement over the competing
black-box attacks varies across the considered SNR range.
}
In addition to improved attack strength, PAMFGA also offers attractive
computational efficiency.
{The case $T=1$ provides no opportunity for inter-task knowledge transfer
and therefore does not exhibit an advantage for PAMFGA: it incurs additional framework overhead and achieves a slightly lower BLER than PAGA. The benefit of PAMFGA becomes evident in the intended multi-surrogate setting.
At $T=4$, PAMFGA achieves the same BLER as PAGA with a lower runtime,
while for $T= 6, 10$, and $14$, it achieves both higher BLER and lower
runtime. These results indicate that the computational and attack-strength
advantages of PAMFGA arise from jointly exploiting multiple related
surrogate tasks rather than from the multifactorial framework itself in the
single-task case.} This reduction stems from sharing a single evolving population across tasks and exploiting implicit genetic transfer, instead of re-running an entire GA from scratch for each surrogate model.

\begin{table}[ht]
    \centering
    \vspace{-.2cm}
   \caption{Running time versus BLER of PAMFGA and PAGA across different numbers of surrogate tasks with $Eb/N0 = 14$.}\label{tab:changingT}
   \vspace{-.1cm}
    \begin{tabular}{ccccccc}
\multicolumn{1}{l}{}                                                     & $T$ & 1 & 4  & 6  & 10 & 14 \\ \hline
\multirow{2}{*}{\begin{tabular}[c]{@{}c@{}}Time\\ (in minutes)\end{tabular}} & PAGA              & \textbf{1.14} & 5.94 & 7.56 & 13.76 & 18.78\\
& PAMFGA            & 1.94 & \textbf{4.25} &  \textbf{5.79} & \textbf{10.32} & \textbf{11.34} \\ \hline
\hline
\multirow{2}{*}{\begin{tabular}[c]{@{}c@{}}BLER\end{tabular}} & PAGA       & \textbf{7e-5} & \textbf{5e-4} & 4e-3 & 0.0086 & 0.0087\\       &  PAMFGA & 6e-5 & \textbf{5e-4} &  \textbf{6e-3} & \textbf{0.0140} & \textbf{0.0140}  \\             \hline
\end{tabular}
\end{table}

Overall, these black-box experiments show that PAMFGA provides a powerful and scalable mechanism for crafting physical adversarial perturbations without any direct access to the target autoencoder. When combined with the strong white-box performance of {PAGA}, the proposed evolutionary framework delivers a unified toolbox for attacking deep-learning-based communication systems under both full-knowledge and limited-knowledge assumptions.

\begin{figure}[pos=t]
    \centering
    \begin{subfigure}[b]{0.48\linewidth}
        \centering
        \includegraphics[width=\linewidth]{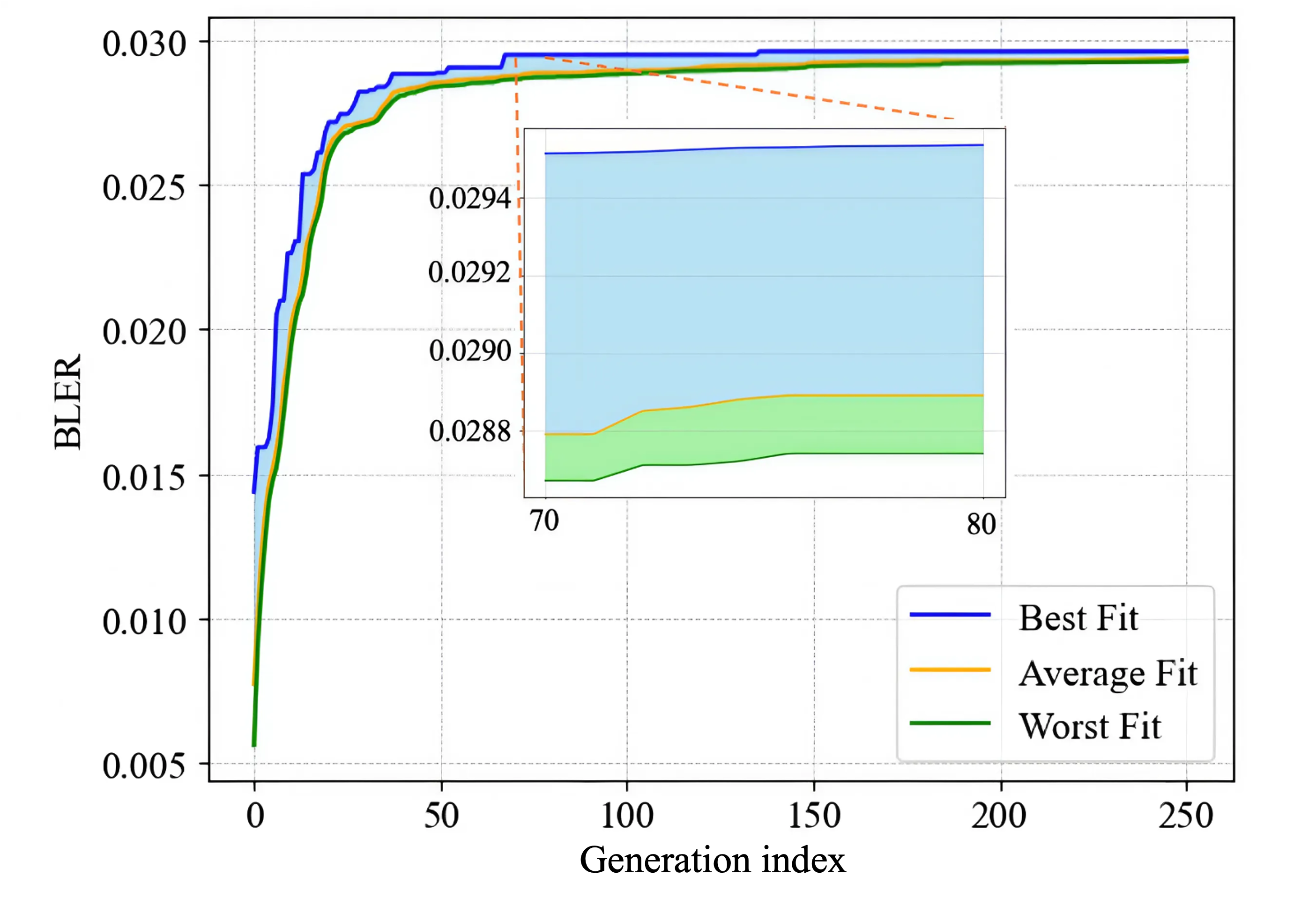}
        \caption{Convergence behaviour of PAGA}
        \label{fig:convergence_6_11}
    \end{subfigure}
    \hfill 
    \begin{subfigure}[b]{0.48\linewidth}
        \centering
        \includegraphics[width=\linewidth]{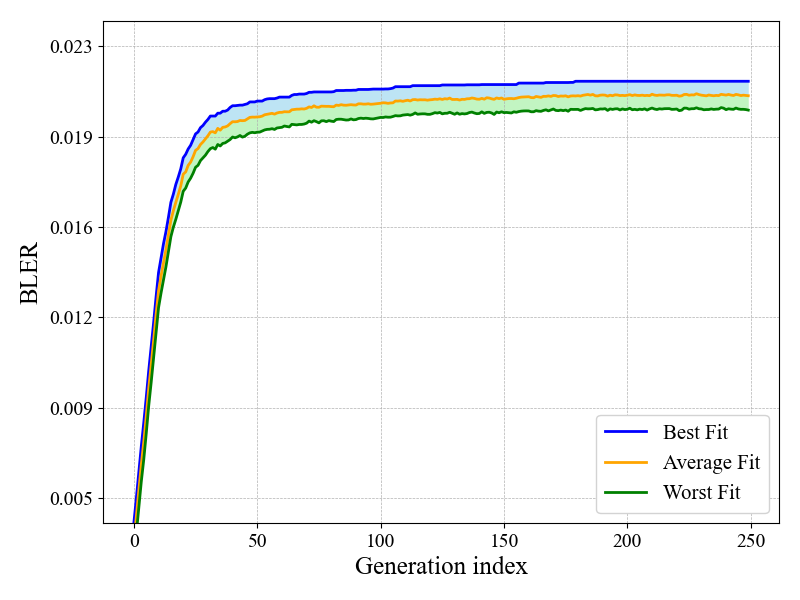} %
        \caption{Convergence behaviour of PAMFGA}  
        \label{fig:second_image}
    \end{subfigure}
    \caption{Convergence behaviour of the proposed algorithms.}
    \label{fig:overall_label}
\end{figure}

{The convergence behaviours of the proposed PAGA and PAMFGA attacks are illustrated in Fig.~\ref{fig:overall_label}, corresponding to the white-box and black-box settings, respectively. In both cases, BLER is used as the fitness metric, where a larger value indicates a stronger adversarial attack. For PAGA [Fig.~\ref{fig:convergence_6_11}], the best, average, and worst fitness values increase rapidly during the first approximately $50$ generations, showing that the population quickly discovers perturbations with substantially improved attack effectiveness. The improvement then becomes progressively smaller, and after about $100$ generations the three curves are nearly flat and tightly clustered. The enlarged view around generations $70$--$80$ further shows that the variation among individuals is only on the order of $10^{-4}$, indicating strong population convergence and stable exploitation of a high-quality region of the search space. A similar convergence pattern can be observed for PAMFGA [Fig.~\ref{fig:second_image}] under the more restrictive black-box setting. Its fitness values also increase sharply during the initial $20$--$40$ generations, followed by a slower improvement phase and a stable plateau after approximately $80$--$100$ generations. Although a small gap remains between the best, average, and worst fitness curves of PAMFGA at convergence, all three curves evolve smoothly and remain stable throughout the later generations, suggesting that the population consistently converges toward high-performing adversarial perturbations despite the absence of direct access to the target autoencoder. Overall, these results demonstrate that the proposed evolutionary search mechanisms exhibit fast and stable convergence under both threat models: PAGA efficiently exploits full white-box information, whereas PAMFGA
achieves similarly reliable search behaviour using surrogate-model fitness
evaluations under the limited-knowledge black-box setting. Together with their respective attack performance, these convergence characteristics support the effectiveness of the proposed evolutionary framework for adversarial attacks under both full-knowledge and limited-knowledge assumptions.}




\paragraph{Robustness under fading channels and practical impairments}
{
The preceding black-box experiments mainly characterize the transferability
and computational behavior of PAMFGA under the default communication setting.
To further examine whether the transferred perturbations remain effective
beyond an idealized AWGN channel, we extend the evaluation to a broader set
of simulated propagation conditions and practical impairments. The considered
settings include Rayleigh fading, Rician fading with different $K$-factors,
channel-state-information (CSI) errors, synchronization mismatch, carrier
frequency offset (CFO), timing offset, and additional path-loss conditions.
We also consider a combined OTA-inspired simulation incorporating multiple
non-ideal effects. These experiments remain simulation-based and are intended
to provide a more realistic robustness assessment rather than to represent
a hardware over-the-air validation.
}

\begin{figure}[pos=t]
    \centering
    \includegraphics[width=\linewidth]{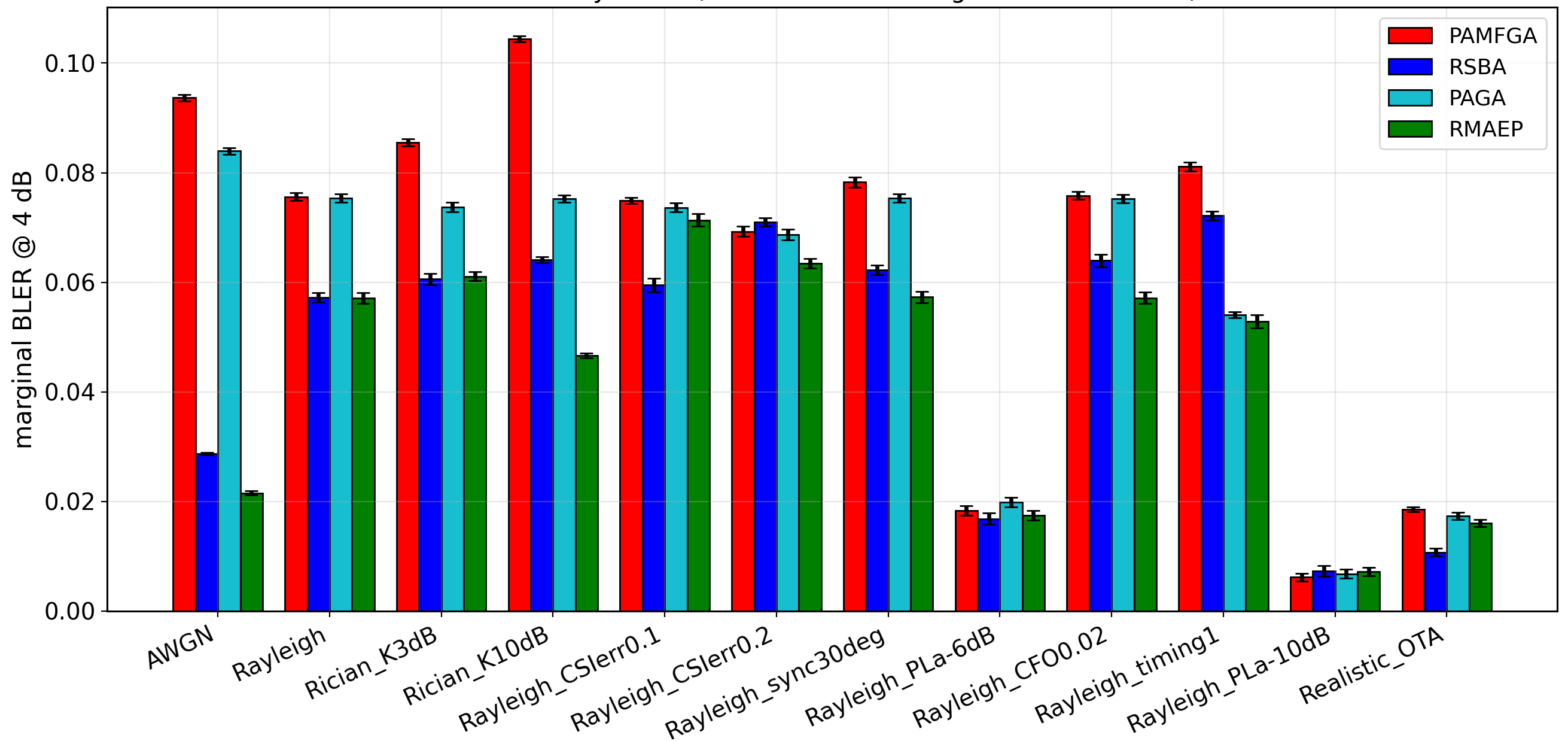}
    \caption{{
    Marginal attack-induced BLER increase at $E_b/N_0=4$ dB under different
    simulated channel and impairment conditions with transmit
    $\mathrm{PSR}=-6$ dB. The marginal BLER is computed relative to the
    corresponding no-attack condition so that the degradation caused by the
    propagation channel itself is removed. Error bars denote 95\% confidence
    intervals over 10 independent runs.}}
    \label{fig:blackbox-channel-effect}
\end{figure}

{
Figure~\ref{fig:blackbox-channel-effect} isolates the degradation attributable
to the adversarial perturbation from that already caused by the underlying
channel. Specifically, the plotted quantity corresponds to the increase in
BLER produced by the attack relative to the no-attack system under the same
channel condition. This distinction is important for fading and impaired
channels, where a high absolute BLER may partly originate from the propagation
environment rather than from the adversarial signal itself. 
The results show that the transferred attacks remain effective across a broad
range of channel conditions. PAMFGA generally produces the largest marginal
BLER increase among the considered black-box attacks, with particularly
visible advantages under AWGN and the considered Rician conditions. Positive
attack-induced degradation also persists under Rayleigh fading, CSI errors,
synchronization mismatch, CFO, timing offset, and additional path loss.
Nevertheless, the magnitude of the attack effect varies substantially with
the propagation condition. In the more severely degraded cases, such as the
stronger path-loss settings, the marginal BLER increase becomes smaller for
all methods. This indicates that channel impairments can attenuate the
effective adversarial perturbation and that attack strength should not be
interpreted independently of the propagation environment.
}

\begin{figure}[pos=t]
    \centering
    \includegraphics[width=\linewidth]{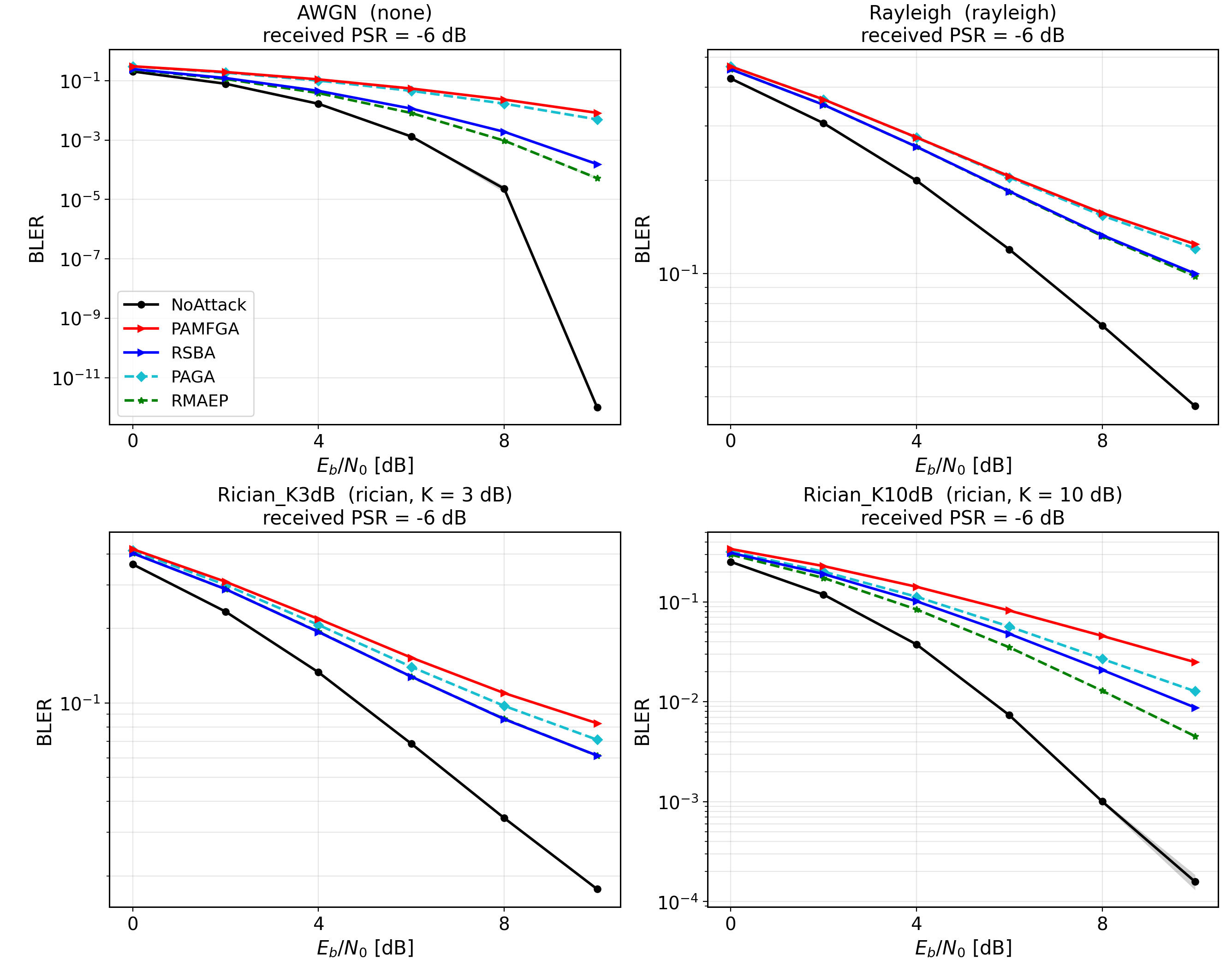}
    \caption{
    BLER versus $E_b/N_0$ for the black-box attacks under representative
    simulated AWGN, Rayleigh, and Rician fading channels with transmit
    $\mathrm{PSR}=-6$ dB. For each propagation scenario, the autoencoder is
    trained end-to-end under the corresponding channel model.}
    \label{fig:blackbox-channel-bler}
\end{figure}

{
To complement the marginal attack-only comparison,
Fig.~\ref{fig:blackbox-channel-bler} reports the absolute BLER over a range of
$E_b/N_0$ values for representative AWGN, Rayleigh, and Rician channels.
Each scenario employs an autoencoder trained under its corresponding channel
model, thereby avoiding an artificial mismatch in which a model trained only
for AWGN is evaluated directly under a different propagation distribution.
The no-attack curve is also reported in each case to provide the channel-specific
reference performance.
}

{
Under AWGN, the adversarial methods substantially increase BLER relative to
the unattacked autoencoder, particularly as $E_b/N_0$ increases and the clean
system approaches a very low error rate. The attack effect also remains
observable under Rayleigh fading, although the absolute BLER of the unattacked
system is already higher because of the more difficult propagation condition.
A similar behavior is observed under Rician fading. 
Among the black-box methods, PAMFGA generally achieves the highest or a
competitive BLER across the representative fading scenarios. The magnitude
of its advantage over RSBA, the SVD-adapted PAGA baseline, and the
SVD-adapted RMAEP baseline depends on the channel and SNR, and the differences
are relatively small at some operating points. We therefore do not interpret
these experiments as evidence that one method uniformly dominates under every
wireless condition. Rather, the main observation is that the adversarial
perturbations produced through surrogate-based transfer remain effective
when the propagation model is changed from AWGN to Rayleigh and Rician fading
and when additional practical impairments are introduced.
}

{
Overall, Figs.~\ref{fig:blackbox-channel-effect} and
\ref{fig:blackbox-channel-bler} provide complementary evidence of robustness.
The first figure separates the attack-induced degradation from the intrinsic
channel-induced error, whereas the second reports the actual end-to-end BLER
experienced by the communication system. Taken together, the results extend
the original AWGN-only evaluation and show that the proposed black-box attack
framework remains effective under a substantially broader range of simulated
wireless conditions, while also revealing that its effectiveness is affected
by fading severity, channel uncertainty, synchronization errors, and path loss.
}

\section{Conclusions}\label{sec:conclusion_future_work}

This paper studied universal physical adversarial perturbations against
end-to-end autoencoder communication systems under white-box and
limited-knowledge black-box settings. We proposed PAGA for white-box attack
generation and PAMFGA for black-box attacks, where multiple substitute
autoencoders are jointly optimized through multifactorial evolution and their
elite perturbations are aggregated for transfer to the unknown target.
The revised experiments show that PAGA generally achieves stronger attacks
than the considered physical-attack and population-based baselines, including
GA, DE, PSO, and CMA-ES under a matched evaluation budget, while its runtime
remains comparable to the population-based solvers. In the black-box setting,
PAMFGA generally achieves the highest or competitive BLER, with gains depending
on the SNR and the number of surrogate models. Transferability is also observed
across different autoencoder architectures and under Rayleigh/Rician fading,
CSI errors, synchronization mismatch, CFO, timing offset, path loss, and
OTA-inspired simulated conditions. These results indicate that high nominal
reliability of learned communication systems does not necessarily imply
adversarial robustness.
The present study remains simulation-based and considers a finite set of
communication configurations, architectures, and surrogate models. Hardware
over-the-air validation, broader MIMO/RIS-assisted and mobile scenarios, and
more diverse target models remain for future work.  Future work will also
investigate adversarial training, perturbation detection, robust autoencoder
design, and joint attack--defense optimization.

\section*{Acknowledgment}
This research is funded by Hanoi University of Science and Technology (HUST) under project number T2025-TN-006.

\subsection*{Declaration of generative AI and AI-assisted technologies in the manuscript preparation process}

During the preparation of this work, the authors used ChatGPT (OpenAI) in order to improve the clarity and readability of the manuscript text. After using this tool/service, the authors reviewed and edited the content as needed and take full responsibility for the content of the published article.

\subsection*{Code availability.} The implementation scripts for PAGA and PAMFGA are publicly available at \url{https://github.com/LuongHvd/PAMFGA}.


\bibliographystyle{elsarticle-num}
\bibliography{references}


\end{document}